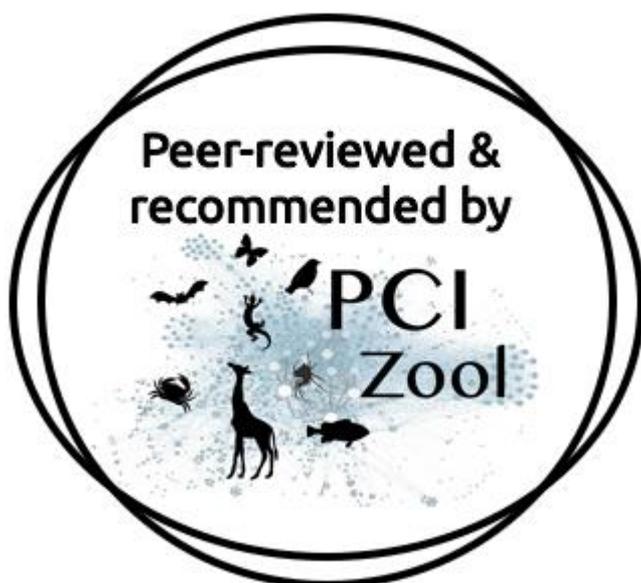

# Revisiting the systematics of Brevipalpus flat mites (Tenuipalpidae): phylogeny, species groups and cryptic diversity

MENDONÇA RENATA SANTOS[1], FERRAGUT FRANCISCO[2], OLIVEIRA ISIS CAROLINA SOUTO DE[3]**, TASSI ALINE DANIELE[4], FELIPE FILENI[5], RONALD OCHOA[6], NAVIA DENISE [7]*

[1]CAPES/PNPD - Faculdade de Agronomia e Medicina Veterinária, Universidade de Brasília, Campus Universitário Darcy Ribeiro, CEP 70.297-400; Embrapa Genetic Resources and Biotechnology CEP 70.770-900, DF, Brasília, Brazil. E-mail mendonca.rsm@gmail.com https://orcid.org/0000-0002-0514-6442

[2]Instituto Agroforestal Mediterráneo, Universitat Politècnica de València, Camino de Vera, s/n, 46022 Valencia, Spain. E-mail fjferrag@eaf.upv.es https://orcid.org/0000-0003-1545-6011

[3]Postgraduate Program in Zoology, Institute of Biological Sciences Universidade de Brasília; Embrapa Genetic Resources and Biotechnology, Cx. Postal 02372, CEP 70.770-900, Brasília, Brazil. E-mail isis.csoliveira@gmail.com https://orcid.org/0000-0001-7690-5269

[4]Escola Superior de Agricultura "Luiz de Queiroz", Universidade de São Paulo, Cx. Postal 9, CEP 13.418-900, Piracicaba, Brazil; University of Florida, Tropical Research and Education Center, Homestead, FL, USA. E-mail alinetassi@gmail.com https://orcid.org/0000-0002-8622-5977

[5]Newcastle University, School of Engineering, Cassie Building, NE1 7RU, Newcastle Upon Tyne, United Kingdom. E-mail felipef93@gmail.com https://orcid.org/0000-0001-7260-1946

[6]Systematic Entomology Laboratory, United States Department of Agriculture, Agricultural Research Service, 20705, Beltsville, Maryland, USA. E-mail ron.ochoa@usda.gov https://orcid.org/0000-0003-1680-3601

[7]*UMR CBGP, INRAE, CIRAD, Institut Agro, IRD, Univ Montpellier, Institut National de Recherche pour l'Agriculture, l'Alimentation et l'Environnement (INRAE), 755 avenue du Campus Agropolis, CS 30016, 34988 Montferrier sur Lez cedex, France. E-mail denise.navia@inrae.fr https://orcid.org/0000-0003-3716-4984

*Corresponding author
Correspondence: E-mail: denise.navia@inrae.fr

**This paper corresponds to part of the Master's research carried out by Isis Carolina Souto de Oliveira and presented to the Postgraduate Program in Zoology at the Institute of Biological Sciences, Department of Zoology, University of Brasília.

**ABSTRACT**

Phytophagous mites in the genus *Brevipalpus* (Tenuipalpidae) can be major pests, causing direct damage to their host plants or vectoring viruses. However, a phylogeny-based classification to understand their evolution and predict their bioecological aspects is lacking. Accurate species identification is crucial for studying pathosystem interactions, implementing quarantine measures, and developing control strategies. This study revisited the classification of *Brevipalpus* based on phylogenetic relationships, identifying cryptic species and determining genetic distance boundaries. A multi-tool exploration of DNA datasets, including mitochondrial (two COI regions) and nuclear (ITS2 and D1-D3) data combined with a detailed morphological study using light and scanning microscopy, was performed. Specimens were collected from 20 host plant families from South America, Europe, and the Middle East. Species were discriminated using three different approaches, namely, Automatic Barcode Gap Discovery (ABGD), Assemble Species by Automatic Partition (ASAP), and a local regression to establish consistent genetic distance thresholds. Results indicate that the current species-group classification only partially matches that based on genetic lineages. Some species currently classified as belonging to the *phoenicis* and *cuneatus* species groups were found to be polyphyletic and related to species placed in other species groups. A rearrangement of the species groups to align with the observed genetic lineages is proposed, and phylogenetically informative morphological traits are defined. Cryptic diversity in certain taxa was consistently confirmed by various methods. Nuclear and mitochondrial markers were essential for identifying candidate lineages when morphology alone was insufficient. Nuclear markers detected host-associated lineages within certain species. Inconsistencies between markers and methods suggest the presence of ongoing speciation processes, and hypotheses about speciation events were explored. Intra- and interspecific genetic distances and threshold values were determined for the four studied genomic fragments and can be used for routine taxonomic identification and uncovering cryptic lineages. Further research should combine genetic data, carefully selected markers, and thorough morphological analyses before proposing new species.

**Keywords:** False spider mites, phytophagous mites, plant virus vector, species-delimitation methods, integrative systematics, molecular taxonomy

## Introduction

Taxonomic classifications based on relationships contribute to the understanding of the evolutionary aspects of organisms, and allow predictions of the biological aspects of a particular taxon based on shared traits with closely related taxa (Gilbert & Webb, 2007; Kellermann et al., 2012). This phylogenetic background can be particularly important for agricultural pests, and is useful for predicting their potential economic impact, their relationship with the phytopathogens they transmit, and for developing control strategies (Kellermann et al., 2012; Godefroid et al., 2016; Jin et al., 2019). However, such background data are still lacking for many important groups of pests, including the phytophagous mites in the genus *Brevipalpus* Donnadieu (Tenuipalpidae).

*Brevipalpus* is one of the most diverse genera in the family Tenuipalpidae and currently includes over 300 valid taxa (Castro et al., 2024). The genus exhibits a broad distribution across most zoogeographical regions, with the Neotropical, Nearctic, and Oriental realms accounting for the majority of described species (Mesa et al., 2009; Castro et al., 2024). *Brevipalpus* mites cause damage to food crops and ornamental plants because they directly feed in large populations on plant tissues and, to a greater extent, they are vectors of plant viruses commonly known as *Brevipalpus* transmitted virus, the citrus leprosis virus (CiLV-C, *Cilevirus leprosis*), and the coffee ringspot virus (CoRSV), which are of economic importance (Kitajima et al., 2010; Childers & Rodrigues, 2011). Many new viruses and *Brevipalpus* species acting as vectors are being discovered (de Lillo et al., 2021), and some of these species are officially regulated as quarantine pests in the international exchange or trade of fresh fruit and propagation material of their host plants (Peña et al., 2015).

The taxonomy of *Brevipalpus* mites has been a challenge for acarologists for decades. Poor descriptions and illustrations as well as lack of a critical study of their morphology have rendered the taxonomy of the genus difficult (Welbourn et al., 2003; Beard et al., 2015). To put this in context, the species previously considered cosmopolitan in distribution and most reported for virus transmission, *Brevipalpus phoenicis* (Geijskes), has been reclassified as *B. yothersi*, and a species complex has been proposed (Beard et al., 2015) to accommodate the multiple species under the name *B. phoenicis*. Additionally, an integrative approach based on DNA sequences of the mitochondrial *cytochrome c oxidase subunit I* (*COI*) and a detailed morphological study has been used to describe and uncover cryptic species in the genus (Navia et al., 2013; Alves et al., 2019).

The genus *Brevipalpus* has variously been divided into species groups by different authors, usually based on just a few morphological traits. Baker *et al.* (1975) and later Baker & Tuttle (1987) proposed six species groups (*cuneatus*, *frankeniae*, *californicus*, *obovatus*, *phoenicis*, and *portalis*) based on the number of lateral setae on the opisthosoma, the number of solenidia (specialized chemosensory setae) on the female tarsus II, and the number of setae on the distal segment of the palps. However, these characters were selected to create an artificial classification specifically designed for the identification of organisms, without any purpose of expressing kinship relationships between them. Morphological studies examining new morphological characters to distinguish the species groups, and DNA analyses, may elucidate these relationships and allow for a redefinition of the species groups based on their phylogenetic relationships.

To date, the small mitochondrial *COI* gene (approximately 400 bp) has been the most frequently used molecular marker in *Brevipalpus* systematic studies (Groot & Breeuwer 2006; Navia *et al.* 2013a; Rodrigues *et al.* 2004; Salinas-Vargas *et al.* 2016; Sánchez-Velázquez *et al.* 2015). However, other markers, such as the nuclear D1-D3 subunit of the 28S gene, have also been employed (Alves et al. 2019). While the COI fragment has been a valuable tool for studying phylogenetic relationships at lower taxonomic levels (see Navia *et al.* 2013), consistent phylogenies require multiple genes. Relying on a single marker reflects only its own evolution and can lead to phylogenetic misinterpretations. Thus, using a broader range of molecular markers, including other mitochondrial and nuclear fragments, may help elucidate the phylogenetic relationships among *Brevipalpus* mites.

As in other groups of organisms, delimiting *Brevipalpus* species can be challenging due to the intraspecific variability in both morphology (Welbourn et al., 2003) and molecular traits (*e.g.,* Navia

*et al.* 2013a). Integrating DNA-based species-delimitation methods with morphological methods can be useful in clarifying taxonomic uncertainties, especially for closely related or recently diverged species (Jorna et al., 2021) or hyper-diverse groups/taxa (Puillandre, Modica, et al., 2012) undergoing pressures of speciation events (Losos & Mahler, 2010; Czekanski-Moir & Rundell, 2019; Shropshire et al., 2020). Furthermore, establishing intra- and interspecific threshold values through highlighting potential new species or cryptic species complexes could aid in the identification of known species, solve taxonomic issues, and play a crucial role in taxonomic initiatives by highlighting potential new species or cases of cryptic species (Hebert et al., 2004).

The main goal of this work was to enhance knowledge of the systematics and phylogenetic relationships within the genus *Brevipalpus* by addressing the following objectives: (**i**) to revisit species-group classification through phylogenetic relationships and evaluate the reliability of the species-group diagnostic characters used in morphological taxonomy; (**ii**) to check for the occurrence of cryptic species within morphologically recognized species by applying multiple species-delimitation approaches and exploring hypotheses regarding speciation mechanisms and; **iii**) to establish species genetic distance boundaries based on polynomial regression analysis.

## Material and methods

### Specimen collection

Specimens of *Brevipalpus* mites were collected between July 2009 to November 2017 from five countries: Argentina, Brazil, Chile, Israel, and Spain. The details of regions of each country from which specimens were collected are presented in Table 1 and Figure SM1-1.

Thirty-three *Brevipalpus* populations were randomly collected from host plants, including fruit trees and ornamental plants, using a 10x hand lens. Samples of leaves and stems from 23 host genera across 20 families of monocotyledon and dicotyledon plants were collected from the field. These samples were transported to the laboratory for further examination under a 40x dissecting stereoscope. Individual mites collected from the same plant and date in a specific location were considered to be part of the same sample.

Specimens of *Brevipalpus* were collected from each sample for morphological studies using light and scanning electron microscopy. Some of the specimens were used for molecular analyses. The specimens used for morphological and molecular analyses were preserved in 70% and 100% ethanol, respectively.

### Morphological identification

For light microscopy specimens preserved in 70% ethanol were directly mounted on microscope slides in Hoyer's medium or in polyvinyl alcohol (PVA) medium, before they were kiln-dried at 56°C for 5–7 days. The slide-mounted specimens were observed under a phase and differential interference contrast microscope (Eclipse 80i Nikon, Tokyo, Japan) at 40x and 100x objectives. Specimens were deposited as vouchers in the plant mite collections at the EMBRAPA Recursos Genéticos e Biotecnologia, Brasilia, Brazil and of Instituto Agroforestal Mediterráneo, Universidad Politécnica de Valencia, Valencia, Spain (codes are listed in **Table 1**).

For scanning electron microscopy, specimens preserved in 100% ethanol were critical point-dried (Leica EM CPD 300), mounted on double-sided carbon tape on stubs, before they were sputter-coated (Baltec SPD 050–Balzers, Lichtenstein). Observations were conducted under a JEOL JSM-IT300 microscope at the Laboratory of Electron Microscopy, ESALQ-USP, in Piracicaba, São Paulo, Brazil.

Morphology-based identification was performed using traditional diagnostic traits, such as the number of solenidia (omega, $\omega$) on tarsus II, the presence of opisthosomal setae pair *f2*, and general dorsal cuticular patterns (Baker 1949; Pritchard & Baker 1956; Baker et al. 1975; Beard at al. 2012; 2015). Additionally, recently valued traits, such as the shape of the spermatheca, type of palp femorogenu seta, detailed dorsal and ventral cuticular patterns, shape of the propodosomal and opisthosomal setae, and leg chaetotaxy, were used (Castagnoli, 1974; Beard et al., 2015).

Identification of certain specimens/populations was confirmed by comparing them with the types of each species deposited in the United States National Museum collection (*B. californicus* Baker, *B. papayensis* Baker, *B. phoenicis*, *B. mallorquensis* Pritchard & Baker) and EMBRAPA Recursos Genéticos e Biotecnologia (*B. incognitus* Ferragut & Navia).

**Molecular study**

*Genomic region selection*

Four target DNA fragments, comprising two mitochondrial and two nuclear, were sequenced to assess the phylogenetic relationships and the genetic variability (intra- and interspecific distances).

The two mitochondrial fragments (**Figure SM1-2**), both in the cytochrome c oxidase I (*COI*) gene were **i)** the 650 bp *COI* corresponding to the "DNA barcoding region" chosen by the Consortium for the Barcode of Life (http://barcoding.si.edu) (Folmer et al., 1994; Hebert et al., 2003), which is among the most widely used barcoding fragments for arthropod groups (Huemer et al., 2014), including Tenuipalpidae (Dowling et al., 2012). The match positions in the mitochondrial genomes of *Brevipalpus yothersi* (Navia et al., 2019) and that of the two-spotted spider mite *Tetranychus urticae* Koch (Tetranychidae) (Van Leeuwen et al., 2008; Grbić et al., 2011) are shown in **Figure SM1-2**; **ii)** the 400 bp (*COI*) selected by Navajas *et al.* (1996), henceforth referred to as *COI* DNF-DNR. It has been previously used in studies of *Brevipalpus* mites with over 300 sequences available in GenBank (Alves *et al.* 2019; Groot & Breeuwer 2006; Navia *et al.* 2013a; Rodrigues *et al.* 2004; Sánchez-Velázquez *et al.* 2015) and has been included to link the results obtained herein with those from previous studies. The two selected nuclear genome fragments were **i)** a 1000-bp portion of the large ribosomal RNA subunit (28S gene) spanning the expansion region D1–D3, henceforth referred to as D1–D3. This region has been used for taxon identification and phylogenetic studies from the species level to higher-level diversification of acariform mites (Dowling et al., 2012; Pepato & Klimov, 2015; Alves et al., 2019), and **ii)** a 480-bp fragment spanning the internal transcribed spacers (ITS), including the partial 5,8S gene and the partial ITS2 region. This region, hereinafter referred to as ITS2, has been used in studies dealing with phylogenies, species diagnostics, and cryptic speciation of Tetranychoidea, the superfamily in which the family Tenuipalpidae is included (Navajas et al., 1999; Mendonça et al., 2011).

*DNA isolation, PCR amplification, and sequencing*

Total genomic DNA was extracted from a single adult female using the DNeasy Tissue kit (Qiagen Germantown, MD, USA) according to the DNA extraction protocol "Purification of Total DNA from Animal Blood or Cells" (SpinColumn Protocol), modified for the extraction of mite DNA (Mendonça et al., 2011; Dowling et al., 2012). For each sample, DNA was extracted from 5 to 20 specimens. The carcasses of most of the specimens recovered from the last stage of the DNA extraction procedure were intact and were subsequently mounted on slides in Hoyer's medium, except for samples from Chile (*Ligustrum* sp. - CH Lisi) and Israel (IS Paed), which contained only a few specimens that could not be retrieved. Recovered specimens were deposited as vouchers in the mite collection of EMBRAPA Genetic Resources and Biotechnology, Brasilia, DF, Brazil.

DNA fragments were PCR-amplified in 25 µL reaction volumes containing *Taq* DNA Polymerase, 2.5 µL of 10× PCR buffer, Standard *Taq* buffer (Qiagen, Brazil), and 2 µL of DNA template. All thermocycling profiles included a final elongation step of 10 min at 72°C. PCR primers and reaction conditions are described in **Tables SM2-1** and **SM2-2**. The primer ITS-F was designed in this study.

PCR products were resolved by electrophoresis on a 1% agarose gel in 0.5X TBE buffer and visualized on Gel Red staining (Biotium, Inc, Hayward, Canada). The amplified fragments containing visible and single bands were directly sequenced on both strands using an ABI 3730XL (Macrogen, Seoul, Korea). No additional primers were used for the sequencing.

**Table 1.** Collection details of the specimens of *Brevipalpus* mites used in the study.

| Morphological identification | Host plant | | Locality | Latitude, Longitude | Collector | Date | Slides (NSPS)* | Voucher code |
|---|---|---|---|---|---|---|---|---|
| | Scientific and common name | Family | | | | | | |
| *B. yothersi* | *Citrus sinensis* (sweet orange) | Rutaceae | Rearing population = genome sequence*** | -15°43'52", -47°54'08" | Navia D., Novelli V.M. | 05.ix.2015 | 5 (5) | BRDF Cisi |
| *B. yothersi* | *Ipomoea batatas* (sweet potato) | Convolvulaceae | Arapiraca, AL, Brazil | -09°45'09", -36°39'40'" | Navia D., Silva E.S. | v. 2013 | 10 (50) | BRAL Ipba |
| *B. yothersi*; *B.* n. *yothersi* 1* *B.* n. *californicus* 1 | *Hibiscus rosa-sinensis* (hibiscus) | Malvaceae | UFRPE, Recife, PE, Brazil | -08°0'53.14", -34°56'59.16" | Gondim Jr M.G. C., Navia D. | 27.vii.2012 | 8 (40) | BRPE Hiro |
| *B. yothersi* *B. incognitus* | *Phoenix* sp. (date palm) | Arecaceae | Piracicaba. SP, Brazil | -22°43'31", -47°38'57" | Kitajima E.W. | 31.v.2013 | 2 (10) | BRSP Phsp |
| *B. yothersi* *B.* n. *yothersi* 2 | *Delonix regia* (flamboyant) | Fabaceae | Piracicaba, SP, Brazil | -22°43'31", -47°38'57" | Kitajima E.W. | 02.vi.2013 | 3 (15) | BRSP Dere |
| *B. yothersi* | *Cecropia pachystachya* (ambay) | Cecropiaceae | Guajará Mirim, RO, Brazil | -10°24'4.96", -65°24'51.94" | Navia D., Ferragut F. | ix. 2012 | 1 (5) | BRRO Cepa |
| *B. yothersi* | *Citrus clementine* (common clementine) | Rutaceae | Bella Vista, Corrientes, Argentina | -34°33'49.34", - 58°41'25.43" | Kitajima E.W. | 26.xi.2013 | 2 (10) | AR Cicl |
| *B. yothersi* *B.* new sp. | *Citrus sinensis* (sweet orange) | Rutaceae | Bella Vista, Corrientes, Argentina | -34°33'49.34", -58°41'25.43" | Kitajima E.W. | 26.xi.2013 | 3 (15) | AR Cisi |
| *B. yothersi* | *Passiflora edulis* (passion fruit) | Passifloraceae | Israel | unknown | Ben David T. | 26.xi.2013 | 1 (5) | IS Paed |
| *B.* n. *yothersi* 2 *B. lewisi* | *Citrus sinensis* (sweet orange) | Rutaceae | Palma del Rio, Cordoba, Spain | +37°41'58.72", -05°16'51.00" | Ferragut F. | x. 2011 | 1 (5) | ES Cisi |
| *B. incognitus* | *Annona muricata* (soursop) | Annonaceae | Bonfim, RR, Brazil | +03°21'36", -59°49'59" | Navia D. | 15.vii.2009 | 8 (40) | BRRR Anmu |
| *B. incognitus* | *Cocos nucifera* (coconut) | Arecaceae | Pacaraima, RR, Brazil | +04°25'52", -61°08'46" | Navia D. | 14.vii.2009 | 4 (20) | BRRR Conu |
| *B.* n. *incognitus* 1 *B.* n. *incognitus* 2 | *Cocos nucifera* (coconut) | Arecaceae | UNIMONTES, Janaúba, MG, Brazil | -15°50'18.60", -43°24'48.11" | Alves R. de B. das N. | ix. 2012 | 10 (50) | BRMG Conu |
| *B. californicus ss*; *B. lewisi* | *Vitis vinifera* (grape) | Vitaceae | Ademuz, Valencia, Spain | +40°03'45", -01°17'04" | Ferragut F., Navia D. | 19.x.2014 | | ESBwA |
| *B. californicus ss*; *B. lewisi* | *Citrus lemon* (lemon) | Rutaceae | Elche, Alicante, Spain | +38°17'28", -00°36'20" | Ferragut F., Navia D. | 20.x.2014 | | ESBcE |
| *B. californicus ss* *B. ferraguti* | *Pittosporum tobira* (japonese cheesewood) | Pittosporaceae | Valencia, Spain | +39°30'3.10", -00°22'15.66" | Ferragut F., Navia D. | 1.x.2011 | 2 (10) | ES Pito |
| *B. lewisi* | *Vitis vinifera spp. sylvestris* (grape) | Vitaceae | Ademuz, Valencia, Spain | +40°03'51", -01°16'49" | Ferragut F., Navia D. | 19.x.2014 | | ESBwC |

| Morphological identification | Host plant | | Locality | Latitude, Longitude | Collector | Date | Slides (NSPS)* | Voucher code |
|---|---|---|---|---|---|---|---|---|
| | Scientific and common name | Family | | | | | | |
| *B. phoenicis* | *Citrus* sp. (orange) | Rutaceae | Ibiuna, SP, Brazil | -23°37'27", -47°19'55" | Kitajima E.W. | 11.iv.2017 | 1(4) | 109_Cisp_I BI_SP |
| *B. papayensis* | *Coffeae arabica* (coffee) | Rubiaceae | Campinas, SP, Brazil | -22°42'30.3", -47°37'53.8" | Tassi A.D. | 24.xi.2017 | 1(1) | 123_Coar_CAM_SP |
| *B. papayensis* | *Coffeae* sp. (coffee) | Rubiaceae | Campinas, SP, Brazil | -22°42'30.3", -47°37'53.8" | Tassi A.D. | 24.xi.2017 | 1(4) | 124_Cosp_CAM_SP |
| *B. papayensis* *B. obovatus* | *Ligustrum sinense* (Chinese privet) | Oleaceae | Brazil, DF, Brasilia | -47°87'00", -15°77'00" | Mendonça, R.S. | 12.i.2006 | | BRDF Lisp |
| *B. ferraguti* | *Tecomaria capensis* (Cape-honeysuckle) | Bignoniaceae | Valencia, Spain | +39°30'3.10", -00°22'15.66" | Ferragut F., Navia D. | 1.x.2011 | 2 (10) | ES Teca |
| *B. ferraguti* | *Myoporum laetum* (ngaio tree) | Scrophulariaceae | Valencia, Spain | +39°30'3.10", -00°22'15.66" | Ferragut F., Navia D. | 1.x.2011 | 2 (10) | ES Myla |
| *B. obovatus* | *Solanum violifolium* ** (creeping violet) | ** Solanaceae | rearing population ESALQ, Piracicaba, SP, Brazil | -22º43'31", -47º38'57" | Kitajima E.W. | 28.v.2013 | 2 (10) | BRSP Sovi |
| *B. obovatus* | *Helichrysum stoechas* (everlasting flower) | Asteraceae | El Saler, Valencia, Spain | +39°22'57.00", -00°19'57.00" | Navia D., Ferragut F. | 13.x.2013 | 5 (25) | ES Hest |
| *B. obovatus* | *Coniza bonariensis* (flaxleaf fleabane) | Asteraceae | El Saler, Valencia, Spain | +39°22'57.00", -00°19'57.00" | Navia D., Ferragut F. | 13.x.2013 | 6 (30) | ES Cobo |
| *B. chilensis* | *Magnolia grandiflora* (majestic beauty) | Magnoliaceae | San Francisco de Mostazal, Cachapoal, Chile | -33°54'58.38", -70°40'31.22" | Trincado R. | 18.ii.2013 | 1 (5) | CHT Mggr |
| *B. chilensis* | *Ligustrum sinense* (Chinese privet) | Oleaceae | Curacavi, Santiago, Chile | -33°28'9.80", -70°43'26.14" | Trincado R. | 28.ii.2013 | 1 (5) | CH Lisi |
| *B. chilensis* | *Magnolia grandiflora* (majestic beauty) | Magnoliaceae | San Francisco de Mostazal, Cachapoal, Chile | -33°54'58.38", -70°40'31.22" | Kitajima E.W. | 08.v.2014 | 1 (5) | CH Mggr |
| *B. chilensis* | *Ribes punctatum* (Chilean currants) | Grossulariaceae | Parque Fray Jorge, Limarí, Chile | -33°37'6.21", -70°42'25.40" | Kitajima E.W. | 08.v.2014 | 1 (5) | CH Ripu |
| *B. mallorquensis* | *Rosmarinus officinalis* (rosemary) | Lamiaceae | Úmbria de los Fresnos (Mijares), Valencia, Spain | +39°28'11.67", -00°22'34.64" | Navia D., Ferragut F. | 12.x.2013 | 2 (10) | ES Roof |
| *B. n. mallorquensis* | *Erica* sp. (heather honey) | Ericaceae | Almedijar, Sierra de Espadan, Castellon, Spain | +39°55'8.18", -00°23'37.22" | Navia D., Ferragut F. | 09.vii.2013 | 6 (30) | ES Ersp |
| *B. oleae* | *Olea europea* (olive tree) | Oleaceae | Jardin del Real, Valencia, Spain | +39°28'50.74", -00°22'3.78" | Navia D., Ferragut F. | 16.x.2013 | 1 (5) | ES Oleu |

* **NSPS** = Number of slides and specimens (in parenthesis) per population/sample; * **n.** = near, putative species disclosed by the diagnostic information
** *Solanum violifolium* [=*Lycianthus asarifolia*]
*** Population from the rearing used to sequencing and assembly the whole-genome of *Brevipalpus yothersi* (Navia *et al.* 2019)

*Dataset and sequence retrieval*

The dataset comprised 399 *Brevipalpus* sequences of the four selected DNA fragments obtained in this study. The distribution of the sequences per fragment is shown in **Table SM2-3**. Sequences were also retrieved from GenBank (332), and their accession numbers and respective authors are shown in **Table SM2-4**. Only *COI* (DNF-DNR) sequences of *Brevipalpus* (328) were available in GenBank, except for one sequence for the subunit D1-D3 (Accession number KP276421) (Pepato & Klimov, 2015) and one from ITS2 (HE984346) (unpublished data). To ensure the accuracy of the dataset, as highlighted by Mendonça *et al.* (2011) for Tetranychidae mites, only sequences from peer-reviewed journals with information on the diagnostic identification methods for the specimens were retrieved from GenBank. Exceptions were made for two unpublished ITS2 sequences, one from *Brevipalpus phoenicis* (HE984346) and the other from *Raoiella macfarlanei* (KP318126), which were included in the datasets after confirmation of their correct assignment. The relationship between these query sequences and their neighboring reference sequences was confirmed using a Neighbor-Joining tree and the Kimura 2-parameter (K2P) distance algorithm (Hebert et al., 2003; Austerlitz et al., 2009). Additionally, a BLAST search in GenBank was conducted to check the accuracy of *COI* (DNF-DNR) and the subunit D1-D3 sequences against other *Brevipalpus* species. Given that the *COI* barcoding region was sequenced for *Brevipalpus* species for the first time in this study, a BLAST search yielded no closely related *COI* barcoding sequences within *Brevipalpus* in GenBank. Instead, the closest matches were from the genus *Raoiella* (Dowling et al., 2012). Also, DNA sequences were obtained in this study for *R. indica* across three gene fragments (*COI* barcoding, *COI* DNF-DNR, and D1-D3) and one fragment (*COI* barcoding) for *Cenopalpus pulcher* Furthermore, *COI* DNF-DNR sequences of *C. pulcher* (GenBank accessions AY320029 and X80873), along with the ITS2 sequence of *R. macfarlanei* (KP318126) retrieved from GenBank, were also included to compose the dataset (**Tables SM2-3 and 4**). The term "sister groups", represented by *C. pulcher*, was applied to taxa considered 'closely related species', while the "outgroup", represented by *Raoiella* sp. (*R. indica* and *R. macfarlanei*), was applied to those considered 'distantly related species', to reflect a gradient of evolutionary relationship among the taxa used for phylogeny reconstruction.

The putative species identified through diagnostic information (morphological traits, phylogenetic analysis, and delimitation methods) received the interim name of its closest species preceded by the prefix "near" and followed by a sequential number in ascending order, for example, *B.* near *yothersi* 1 and *B.* near *yothersi* 2 or simply *B.* near *mallorquensis*, abbreviated as *B.* n. *yothersi* 1, and so on. The new species in branches without closely related taxa were designated as *Brevipalpus* new sp (*B.* new sp.). Further taxonomic studies presenting complete species descriptions will be conducted to assign scientific names to these putative new taxa.

*Sequence edition and alignments*

The Staden Package v.1.6.0 (Staden et al., 1998) was used for checking the quality, editing, and assembling the raw data into sequence contig. The DNA sequences were aligned by the ClustalW multiple alignment procedure using BIOEDIT 7.0 (Hall, 1999) and Muscle (Muscle: multiple sequence alignment comparison by log–expectation program) (Edgar, 2004). No manual adjustments of the alignments were made. The alignment of *COI* sequences was confirmed by translating the aligned DNA into amino acids using the Expasy server, https://web.expasy.org/translate/ (Gasteiger, 2003). To identify candidate protein-coding regions in DNA sequences, an open reading frame was determined using the graphical analysis tool (ORF FINDER) available at https://www.ncbi.nlm.nih.gov/orffinder/. The Smart BLAST tool in the National Center for Biotechnology Information (https://blast.ncbi.nlm.nih.gov/smartblast/smartBlast.cgi) was used to detect aberrant or unusual sequences.

Nucleotide composition (calculated as the base frequencies for each sequence and an overall average), substitution patterns, and rates were estimated using jModeltest version 2.1.10 20160303 (Darriba et al., 2012) according to the selected model (**Table SM2-5**). The homogeneity of the substitution patterns between sequences was tested with the Disparity Index (ID).

**Phylogenetic analysis**

The phylogenetic relationship among taxa was established using Maximum Likelihood (ML), Neighbor Joining (NJ), and Bayesian Inference (BI). For ML analyses, jModeltest was used to estimate the best-fit nucleotide substitution models using the Akaike information criterion corrected (AICc) and the Bayesian information criterion for each dataset. The proportion of invariable sites (I) and gamma-distributed rates (G) defined in jModeltest were conserved in all models. The best-fit models as well as information on the ML parameters are shown in **Table SM2-5**.

The ML analyses were performed using the online version of PhyML v. 3.0 (Guindon & Gascuel, 2003; Guindon et al., 2010), and NJ analysis was performed using MEGA 7.0 (Kumar et al., 2016). NJ trees were constructed by treating gaps as missing data. The Bayesian Information analysis was performed using MrBayes V.3.2.6 (Ronquist et al., 2012). The topologies of the trees obtained using ML, NJ, and BI were compared. The robustness of the trees was assessed with a bootstrap analysis with 1,000 replicates. Topologies with posterior probabilities of 0.95 were regarded as well supported (Wilcox et al., 2002).

The distributions and frequencies of haplotypes (*COI* fragments) and sequence variants (genotypes) (D1–D3 and ITS2) were inferred using DnaSP version 6 (Rozas et al., 2017). When identical sequences were found in the alignment, a single sequence of each group was included to produce the ML tree, and the number of times the identical sequences occurred in the dataset was indicated in brackets in the phylogenies.

*Bayesian Combined Analysis*

To perform the Bayesian Combined Analysis (BCA), unique variants (haplotypes/genotypes) of the nucleotide sequences of each fragment were concatenated according to the *Brevipalpus* clusters observed in the output files of the ML individual gene tree analyses. Any *Brevipalpus* specimen for which fewer than two fragments were sequenced was excluded.

The concatenated sequences were achieved for three sets: i) a complete DNA set, including the four fragments (two mitochondrial and two nuclear); ii) the two mitochondrial *COI* fragments, and iii) the two nuclear fragments. The sequence files were individually organized using MEGA v.7. Alignment of the four fragments was performed separately with Muscle v3.8.31. The files containing 19 *Brevipalpus* taxa were concatenated in three matrices (complete, mitochondrial, and nuclear) in Mesquite v.3.0.4 (Maddison & Maddison, 2021). The complete set alignment had 74 sequences totaling 2,653 base pairs (bps) (*COI* barcoding = 661 bps; *COI* DNF–DNR = 369 bps; D1–D3 = 1,001 bps; ITS2 = 622 bps), and the mitochondrial and nuclear set alignments had 60 sequences/1030 bp and 51 sequences/1623bp, respectively.

The BCA was performed in MrBayes v.3.2. The number of categories used to approximate the gamma distribution was set at four, and four Markov chains were run for 10,000,000 generations; the final average standard deviation of the split frequencies was less than 0.01, and the stabilization of the model parameters (burn-in = 0.25) occurred at approximately 250 generations. The phylogenetic trees, based on the MrBayes output file (Newick format) created by the PhyML 3.0 algorithm and MrBayes program, were edited using FigTree v.1.4.3 (http://tree.bio.ed.ac.uk/software/figtree/) (Rambaut, 2009). A FASTA-format file of DNA sequences is available on the Zenodo repository (see Appendices files).

*Assessing genetic distances to support species identification*

Analyses of the overall pairwise genetic distances (distribution and frequencies) and between nucleotide sequences were performed for each DNA fragment using MEGA v. 7 (Kumar et al., 2016). The analyses of intra- and interspecific divergences were explored among the taxa assignments based on multiple methods, including: **i)** distance threshold-based approach to elaborate distance matrices between intra- and interspecific distances; **ii)** DNA-based species-delimitation using Automatic Barcode Gap Discovery (ABGD) (Puillandre, Lambert, et al., 2012); **iii)** Assemble Species by Automatic Partitioning (ASAP) (Puillandre et al., 2021), and **iv**) intra- and interspecific thresholds based on polynomial regression fitting.

The genetic distances were analyzed and compared with the observed morphological divergences between close and putative species, searching for a link between these three approaches (genetic distance, phylogeny, and morphology). The distance matrices were constructed using the Kimura 2-parameter model (K2P) (Kimura, 1980) and uncorrected *p*-distance, and the standard error estimates were obtained using a bootstrap procedure (1,000 replicates) in MEGA v.7. The K2P model was used to make these results suitable for comparison with previous studies on Acari, such as Tetranychidae (Mendonça et al., 2011; Sakamoto et al., 2017), Eriophyidae (Skoracka et al., 2012) and Phytoseiidae (Lima et al., 2018). Genetic distances are usually calculated without models selected for specific data (Strimmer *et al.* 2009; and Srivathsan & Meier, 2012). A simple distance measure was implemented in order to allow for comparison with other diversity studies (Collins et al., 2012). Therefore, the uncorrected p-distances with SE estimates (bootstrap procedure 1,000 replicates) were also accessed using MEGA 7.

*Species hypotheses based on intra- and inter-diversities for closely related taxa*

The intra- and interspecific genetic distances (K2P and *p*-distances) between *Brevipalpus* species, including the putative and new species, were calculated to explore the values that separate one taxon from another. Scenarios and hypotheses were evaluated to estimate the distance values, *i.e.,* the genetic distances for various taxa combinations, considering: **i)** the morphological similarities and **ii)** the architecture revealed by the phylogenetic tree topologies obtained through the BCA (**Figures 1, 2A, and 2B**), as well as through the ML phylogenies performed separately for each fragment (**Figures SM1-3–SM1-6).** There was a convergence of the analyzed hypotheses and the clades revealed by phylogeny; therefore, only the results related to the clusters revealed by the concatenated phylogeny will be discussed.

The distance matrices within and between *Brevipalpus* species were processed as follows: **i)** mean distance values; **ii)** the maximum and minimum values observed within the same species (intraspecific variation), and **iii)** the maximum and minimum values observed between two species (interspecific variation). Minimal interspecific variation is assumed to be the Nearest Neighbor according to Kekkonen *et al.* (2015). For each fragment, the lowest reliable genetic divergences between taxa with clearly different taxonomic statuses, *e.g., B. chilensis* versus *B. obovatus*, were used to guide assumptions regarding the taxonomic status of the putative species, as indicated by the combined phylogenies and morphological observations (Navia et al., 2013).

*DNA-sequence-based species-delimitation method*

The ABGD algorithm was used to determine the occurrence of various species within a dataset based on the gap between clusters of sequence pairs (Puillandre, Lambert, et al., 2012) on the web interface (https://bioinfo.mnhn.fr/abi/public/abgd/abgdweb.html). To investigate a range of partitions from conservative to liberal species-delimitation schemes, three values of the relative gap width (X = 1.0, 1.25, 1.5) were tested, and two distance metrics were used (K2P and *p*-distance). For the range of prior intraspecific divergences (P), the sampled values were estimated from $P_{min}$ = 0.001 to $P_{max}$ = 0.1. Defaults were used for all other parameter values. Prior intraspecific divergence (*P*) ranged from 0.001 to 0.1, with steps set to 10 and Nr bins (for distance distribution) set to 20. Parameters for ABGD were analyzed to select those whose results better matched known species. ABDG analyses were performed per *Brevipalpus* species that were clustered in clades identified by phylogeny.

The ASAP (Puillandre et al., 2021) was used to detect the hypothesized partitions of the species and the *Brevipalpus* species boundaries. Multiple sequence alignments were analyzed using ASAP Web Server (https://bioinfo.mnhn.fr/abi/public/asap/asapweb.html), with the "ASAP-score" and the probability value considered to select the optimal number of species partitions, as well as the partitions within the range of genetic distances of 0.005 and 0.05, regarding the K2P and simple distance *(p*-distances). The ASAP analysis was performed using the complete set of sequences per fragment.

*Intra- and interspecific thresholds based on the polynomial regression fitting*

Intra- and interspecific genetic divergences for the four DNA fragments were examined for a finer characterization of the levels of divergence among species (Robinson et al., 2009). The analyses were elaborated using both the formal species, putative species, and new species (19 taxa) to define reliable threshold values to discriminate taxa. Two distance types were computed: the maximum intradifference observed within the same species (maximal intraspecific variation) and the minimum interdifference observed between two species of the same genus (minimal interspecific variation), assumed to be the Nearest Neighbor (Kekkonen et al., 2015) using the species selected from the database. Intradifferences were sorted in descending order whereas interdifferences were sorted in ascending order. The position of each point was divided by the sample size, resulting in two curves with values distributed between 0 and 1 for comparison. For instance, the *COI* barcoding region had 136 interdifference values that were ordered in descending order. The minimum difference between *B. oleae* and *B. ferraguti* was 18.2%, and was the thirteenth largest difference among species; therefore, the *B. oleae-B. ferraguti* difference was represented by the point: x = 13/136 = 0.095 and y = 0.182. A margin of error for each point was obtained based on the standard error of the sample and a significant threshold of α <0.05.

The focus was on the intersection between the curves of the intra- and interspecific values and the position where the interspecific values were higher than the intraspecific ones. As both curves were modeled with points that presented a margin of error instead of a point of intersection, the curves have an ambiguous zone where the intra- and interspecific values cannot be differentiated at the previously defined significance level (**Figure SM1-7**). The modeling of intra- and interspecific values was performed with a local polynomial regression fitting (Savitzky & Golay, 1964). Use of nonlocal regressions, such as linear, exponential, and logarithmic regression, resulted in a goodness-of-fit that was often less than 0.8. The analysis presents a convergence area, as a polygon, where the closest species are mixed (**Figure SM1-7**). Within this area, it is challenging to delineate species based solely on genetic distance criteria, as some intraspecific values overlap or even match up with interspecific values, violating the gap concept of the barcoding premise (Hebert et al., 2003; Meyer & Paulay, 2005). The overlap of these values could represent incompletely sampled groups, particularly when the coalescent process has not fully sorted between incipient species (ancestral polymorphism), giving rise to genetically polyphyletic or paraphyletic species (Rosenberg, 2003). The intersection point (*y*) in the convergence area represents the average value where the intraspecific distance meets the interspecific distance. Values that escape the gap (where differences within a species reach high values exceeding the minimum value found between two different species, *α*) were regarded as *threshold guide values* to drive diagnostic studies. Genetic distance values within the confidence interval suggest possible co-specificity, whereas values above the upper limit indicate different taxa. The threshold-based approach was applied to indicate a reliable distance interval where there was no overlap of *Brevipalpus* intra- and interspecific distance values, especially when dealing with a large set of DNA sequences.

# Results

**Datasets, haplotypes, and sequence diversity indexes**

The datasets of the four DNA fragments consisted of 2,653 base pairs (bp) of nucleotides, with 1,030 bp from the mitochondrial *COI* barcoding region ($\cong$650 bp), *COI* DNF–DNR ($\cong$364 bp), and 1,623 bp from the nuclear ribosomal regions, including subunit D1–D3 ($\cong$1,000 bp) and ITS2 ($\cong$480 bp). The number of sequences analyzed varied between fragments, with the most frequent being those of *COI* DNR–DNF (432 sequences), followed by those of ITS2 (118), D1–D3 (97), and *COI* barcoding (84) (**Table SM2–3**).

The assigned species names, DNA sequence variation (haplotypes, genotypes), number of times sequences occurred in the dataset (frequency), and respective GenBank accessions are shown for each fragment in **Tables SM2–6 - 9** for *COI* barcoding, *COI* DNF-DNR, subunit D1-D3, and ITS2, respectively.

Intraspecific comparison of genetic diversity indices was estimated for each fragment (**Table SM2-10**). The number of variable sites, the average number of differences in nucleotides (K), and the diversity of nucleotides (*Pi*) showed high variability in the mitochondrial and nuclear sequences. Nucleotide diversity (*Pi*) ranged from 5.52% to 9.09% and from 6.24% to 12.58% for the mitochondrial and nuclear fragments, respectively. The average number of differences in nucleotides (K) and nucleotide diversity (*Pi*) showed high variability in the mitochondrial and nuclear sequences of *Brevipalpus* taxa.

**Molecular phylogeny of *the Brevipalpus* species**

The general topologies of the phylogenetic trees inferred by the NJ, ML analyses, and BI were congruent and revealed a similar pattern for all four fragments. Thus, only the ML phylogeny is shown for each fragment. The BCA phylogeny-based on the complete DNA dataset produced a well-supported tree (**Figure 1**). Nodes representing the closest relationships had a Bayesian posterior probability (BPP) of one, and were also consistently recovered by the ML methods for each fragment individually (**Figures SM1-3 - 6**).

Phylogenetic analysis grouped the *Brevipalpus* species/putative species into six main clades: **I**-*Brevipalpus yothersi*, *B.* near *yothersi* 1, *B.* near *yothersi* 2, *B. incognitus*, *B.* near *incognitus* 1, and *B.* near *incognitus* 2; **II**-*Brevipalpus* n. sp., composed of only specimens collected from *Citrus* from Argentina; **III**-*B. californicus s.s.* (according to Beard et al. 2012), *B.* near *californicus* 1, *B.* near *californicus* 2, and *B. lewisi*; **IV**-*B. phoenicis*, *B. papayensis*, *B. ferraguti*, *B. obovatus*, and *B. chilensis;* **V**-*B. mallorquensis* and *B.* near *mallorquensis,* and **VI**-consisting of only *B. oleae*. One of the deepest nodes clustered clades I, II, III, and IV, while another one clustered clades V and VI. The species in Clade I remained separated from the other three groups (II, III, and IV). Clades III and IV grouped together from a common ancestor.

The tree based on the concatenated mitochondrial sequences (**Figure 2A**) generally agrees with the tree based on the complete DNA set. Most nodes were highly supported, except for the clustering of clades III and IV, which had a support value of BPP = 0.81 (compared to BPP = 1 for the complete combined tree). For the tree based on the concatenated nuclear regions (**Figure 2B**), most of the basal nodes were well supported (BPP ≥ 0.99). However, the tree based on the concatenated nuclear regions was different in structure from the one based on the complete dataset in that Clade II was nested with Clade I making a weakly supported node (BPP = 0.61). In contrast, clades III and IV both showed high support (BPP = 1) in the complete DNA dataset and concatenated *COI*.

The species and putative species and the hypothesized new species shown by the phylogenies (CBA and ML) obtained for each fragment are summarized in **Table 2**. The taxa observed throughout the phylogenies were congruent and converged to a similar number of assigned species, taking into account the missing sequences. Eleven species were recorded for the *COI* fragments and the subunit D1-D3. For the ITS2 region, the dataset included ten species, as no sequences of *B. oleae* were obtained. The number of putative species uncovered by the phylogenies differed among the fragments due to missing sequence data for some taxa **(Table 2)**.

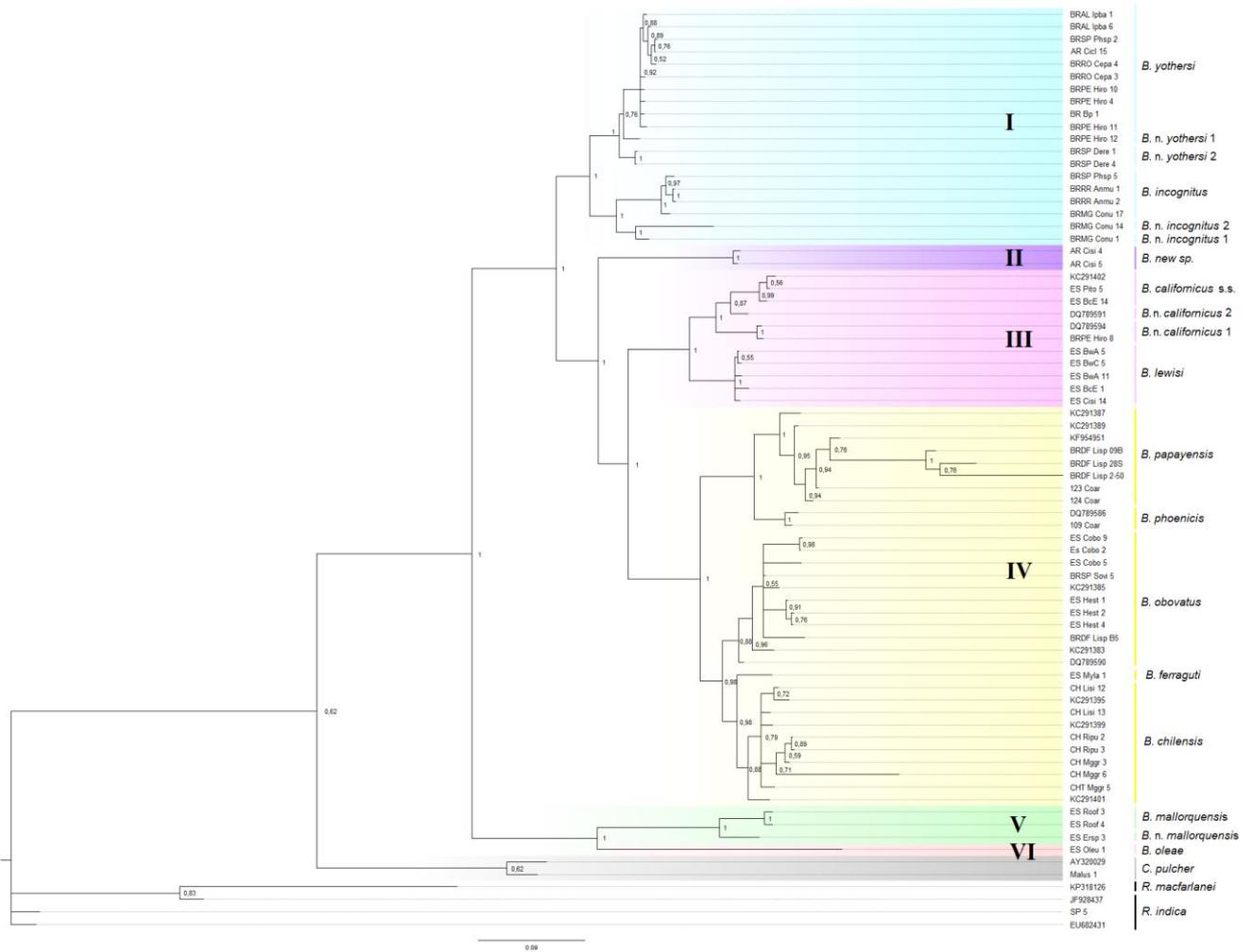

**Figure 1.** Combined Bayesian inference (BI) tree analysis for *Brevipalpus* species reconstructed from four genomic fragments: the mitochondrial *cytochrome c oxidase subunit I* (*COI*) (*COI* barcoding region and *COI* DNF–DNR), the ribosomal subunit D1–D3 in the 28S gene (D23F & D6R primers), and the internal transcribed spacer II (ITS2) sequences. Bayesian posterior probability values are shown. Only probabilities greater than 0.5 (>0.5) are indicated above the branches. The species names based on morphological identification are to the right of the voucher code (**Table 1**). Candidate species were numbered in ascending order after the name of their closest species. *Brevipalpus* species groupings are highlighted in colored squares: blue = Clade I; purple = Clade II, hitherto with a single representative species (*B.* n. sp from Argentina); rose = Clade III; yellow = Clade IV; green = Clade V and light rose = Clade VI. *Cenopalpus pulcher* and *Raoiella* sp. (*R. indica* and *R. macfarlanei*) were used as sisters and outgroups, respectively.

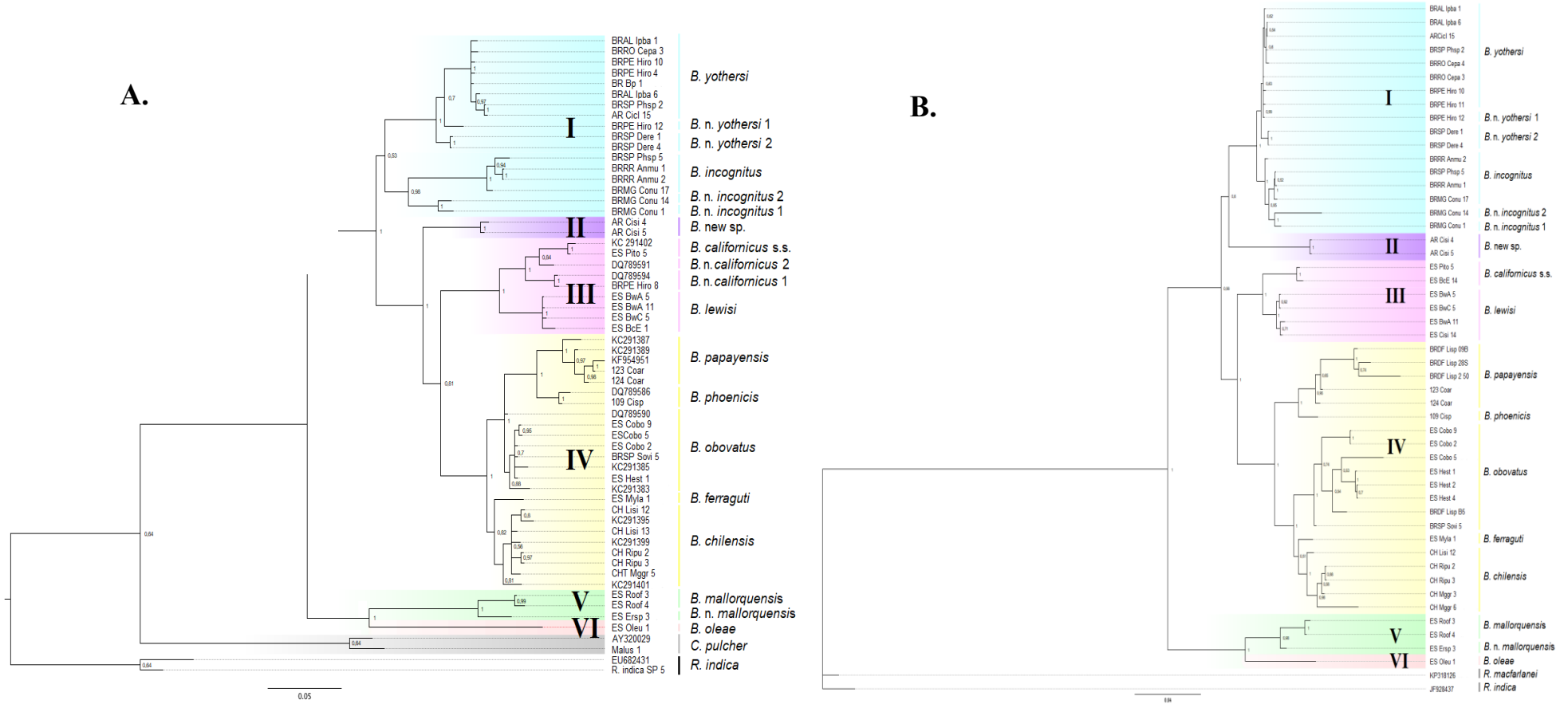

**Figure 2.** Bayesian phylogeny constructed using sequence data obtained from the concatenated dataset consisting of: A. mitochondrial fragments (*COI* barcoding region and *COI* DNF–DNR); B. nuclear fragments (subunit D1–D3 28S and ITS2). Statistical supports indicate the Bayesian posterior probability values. Only probabilities greater than 0.5 (>0.5) are indicated above the branches. The species names based on morphological identification are to the right of the voucher code (**Table 1**). Candidate species were numbered in ascending order after the name of the closest species. Putative *Brevipalpus* species groupings are highlighted in colored squares: blue = Clade I; purple = Clade II, hitherto with a single representative species (*B.* n. sp from Argentina); rose = Clade III; yellow = Clade IV; green = Clade V and light rose = Clade VI. *Cenopalpus pulcher* and *Raoiella* spp. (*R. indica* and *R. macfarlanei*) were used as sisters and outgroups, respectively.

**Table 2.** Sets of DNA sequences of *Brevipalpus* species and putative species supported by phylogeny for each DNA fragment studied. Gray blocks indicate taxa with nonamplified DNA sequences for the respective fragments.

| Clade | *COI* barcode | *COI* DNF-DNF | subunit D1-D3 | ITS2 |
|---|---|---|---|---|
| Clade I | *B. yothersi* | *B. yothersi* | *B. yothersi* | *B. yothersi* |
| | *B.* near *yothersi* 1 | *B.* near *yothersi* 1 | *B.* near *yothersi* 1 | - |
| | *B.* near *yothersi* 2 | *B.* near *yothersi* 2 | *B.* near *yothersi* 2 | *B.* near *yothersi* 2 |
| | *B. incognitus* | *B. incognitus* | *B. incognitus* | *B. incognitus* |
| | *B.* near *incognitus* 1 | - | *B.* near *incognitus* 1 | *B.* near *incognitus* 1 |
| | *B.* near *incognitus* 2 | *B.* near *incognitus* 2 | *B.* near *incognitus* 2 | *B.* near *incognitus* 2 |
| Clade II | *B.* new sp. | *B.* new sp. | *B.* new sp. | *B.* new *sp.* |
| Clade III | *B. californicus s.s.* | *B. californicus s.s.* | *B. californicus s.s.* | *B. californicus s.s.* |
| | *B.* near *californicus* 1 | *B.* near *californicus* 1 | - | - |
| | - | *B.* near *californicus* 2 | - | - |
| | *B. lewisi* | *B. lewisi* | *B. lewisi* | *B. lewisi* |
| Clade IV | *B. phoenicis* | *B. phoenicis* | *B. phoenicis* | *B. phoenicis* |
| | *B. papayensis* | *B. papayensis* | *B. papayensis* | *B. papayensis* |
| | *B. ferraguti* | *B. ferraguti* | *B. ferraguti* | *B. ferraguti* |
| | *B. obovatus* | *B. obovatus* | *B. obovatus* | *B. obovatus* |
| | *B. chilensis* | *B. chilensis* | *B. chilensis* | *B. chilensis* |
| Clade V | *B. mallorquensis* | *B. mallorquensis* | *B. mallorquensis* | *B. mallorquensis* |
| | - | *B.* near *mallorquensis* | - | *B.* near *mallorquensis* |
| Clade VI | *B. oleae.* | *B. oleae.* | *B. oleae.* | - |

**Genetic distance-based analysis**

*Species hypothesis using intra- and intergenetic distances*

Distance values for overall sequence pairs for each genome fragment, as well as exclusively for sequences of *Brevipalpus* species (K2P and *p*-distance), are summarized in **Table 3**. The mean intraspecific variability of the studied *Brevipalpus* species/putative species is also presented. The highest mean variabilities were recorded in the mitochondrial *COI* barcoding and the nuclear ITS2. Single matrices of distances of pairwise comparisons between and within *Brevipalpus* species/putative species for all fragments are presented in **Tables SM2-11**–**SM2-18**.

**Table 3.** Mean overall interspecific divergence for the sequence pairs and for sequences of *Brevipalpus* species, minimum and maximum divergence, and the standard error (SE). The mean intraspecific variability (minimum and maximum) observed in *Brevipalpus* species. Values indicate Kimura 2 parameters and *p*-distance.

| Genomic region | Interspecific divergence | | | | Intraspecific divergence | |
|---|---|---|---|---|---|---|
| | K2P distance Mean (%) range (%) (S.E.%) | | *P*-distance Mean (%) range (%) (S.E.%) | | *Brevipalpus* Mean (%) (min.-max.%) | |
| | All sequences | *Brevipalpus* sequences | All sequences | *Brevipalpus* sequences | K2P dist. | *p*-distance |
| *COI* barcoding | 11.31 (0.0-27.7) (1.56) | 9.86 (0.0-21.86) (1.44) | 10.28 (0.0-23.09) (1.29) | 9.11 (0.0–18.95) (1.22) | 1.12 (0.0–**2.44**) | 1.11 (0.0–**2,40**) |
| *COI* DNR-DNF | 5.55 (0.0-21.78) (1.09) | 5.45 (0.0–18.77) (1.19) | 5.23 (0.0–18.87) (1.10) | 5.14 (0.0–16.60) (1.09) | 0.91 (0.0–**4.28**) | 0.79 (0.0–**4.15**) |
| D1–D3 | 7.13 (0.0-33.41) (1.1) | 6.65 (0.0-22.70) (1.06) | 6.48 (0.0–26.45) (0.95) | 6.11 (0.0–19.24) (0.93) | 1.90 (0.0–**2.86**) | 1.87 (0.2–**2.81**) |
| ITS2 | 12.11 (0.0–93.52) (2.32) | 11.00 (0.0–42.02) (2.19) | 10.42 (0.0–52.07) (1.75) | 9.79 (0.0-31.80) (1.72) | 4.83 (0.0–**13.07**) | 4.60 (0.0–**11.98**) |

According to the COI barcoding region, the lowest interspecific genetic divergences were recorded between: i) *B. chilensis* and *B. ferraguti* ($\bar{X}$ distance = 3.11%, ranging from 2.64% to 3.49%); and ii) *B. obovatus* and *B. chilensis* ($\bar{X}$ = 3.27%, 2.83–3.70%) (Tables SM2-11–SM2-18). Because these species are recognized as the most closely related taxa (Navia et al. 2013), these reference values suggest that *B.* near *yothersi* 1 and *B.* n. *yothersi* 2 represent a taxon distinct from *B. yothersi* ($\bar{X}$ = 3.75% and 3.81% distances, respectively). However, these two putative taxa could not be discriminated from each other ($\bar{X}$ = 2.18%, 1.96%–2.40%) because the genetic divergence values were lower than those in the references. The same pattern was observed for *B. n. incognitus* 1 and *B. n. incognitus* 2, *i.e.*, these putative taxa can be regarded as different from *B. incognitus* ($\bar{X}$ = 7.66 and 7.44% distances, respectively), but the reference values did not support a partition between each other ($\bar{X}$ = 1.99%). The COI barcoding region robustly discriminated *B.* n. *californicus* 1 from *B. californicus s.s.* ($\bar{X}$ = 4.58%), indicating that they are two different species."

The *COI* DNF-DNR distance values were lower overall than those observed for the barcoding fragment. The lowest interspecific genetic divergence for the *COI* DNF–DNR fragment was observed between *B. chilensis* and *B. obovatus* ($\bar{X}$ = 3.25%, 1.91%–4.70%), similar to that observed in the *COI* barcoding region. These values also indicate that the putative species *B.* n. *incognitus* 2 is differentiated from *B. incognitus* by a large distance ($\bar{X}$ = 6.59%, 5.90–7.14%). *Brevipalpus californicus s.s.* was separated from *B.* near *californicus* 1 ($\bar{X}$ = 3.57%, 3.48%–4.29%) and from *B.* near *californicus* 2 ($\bar{X}$ = 2.70%, 2.70%–2.70%). However, the distance between *B.* near *californicus* 1 and *B.* n. *californicus* 2 ($\bar{X}$ = 2.37%, 2.3–3.08%) was not enough to discriminate as separate species. *Brevipalpus mallorquensis* was differentiated from *B.* n. *mallorquensis* ($\bar{X}$ = 3.69%, 3.49–3.89%) (**Tables SM2-11–SM2-18**).

The lowest D1–D3 interspecific distance was recorded between *B. yothersi* and *B. incognitus* ($\bar{X}$ = 0.68%, 0.40%–1.01%) and between *B. phoenicis s.s.* and *B. papayensis* ($\bar{X}$ = 1.52%, 1.21–1.83%) (**Tables SM2-11–SM2-18**). Based on these lowest reference values, the putative species *B.* n. *incognitus* 1 could be separated from *B.* n. *incognitus* 2 ($\bar{X}$ = 1.77%, 0.81%–3.49%). These two putative species also showed distance values that clearly distinguished them from *B. incognitus-B. incognitus* x *B.* n. *incognitus* 1 ($\bar{X}$ = 1.6%, 0.40%–3.49%) and *B. incognitus* x *B.* n. *incognitus* 2 ($\bar{X}$ = 1.18%, 1.011–1.21%). Even according to these low reference values, *B. yothersi* could not be differentiated from its related putative species and not even *B.* n. *yothersi* 1 x *B.* n. *yothersi* 2. All other putative species could be discriminated by D1–D3 distances.

For the nuclear ITS2 fragment, B. *californicus* ss. and *B. lewisi* exhibited the lowest interspecific distance ($\bar{X}$ *distance* = 0.93%, 0.93%–1.40%). All the putative species delimited by morphology and phylogeny, *i.e.*, *B.* n. *yothersi* 2, *B.* n. *incognitus* 1, *B.* n. *incognitus* 2, *and B.* n. *mallorquensis*, could be discriminated by the output values of the ITS2 distances.

### *ABGD-Automatic Barcode Gap Discovery for species-delimitation method*

The ABDG results, including relative gap width (X), prior intraspecific distance (P), prior maximal distances (PMD), and barcode gap distance (BGD) computed by the K2P and *p*-distance model algorithms, are shown in **Table 4**. The number of candidate species was consistent whether analyzing the total set of sequences or when separated by clades, therefore, only the results from the total set were discussed.

Analysis of the *COI* barcoding dataset using the K2P distance model recovered 20–21 Molecular Operational Taxonomic Units (MOTU's, Floyd *et al.* 2002) (**Table 4)**. The 17 species identified through phylogeny/morphology and three additional taxa within the *B. chilensis* taxon— specimens from *Ribes punctatum* (MG458824), *Magnolia grandiflora* (MG458832), and *Ligustrum sinense* (MG458834) - were recognized across various relative gap widths (X = 1.0, 1.25, and 1.5). Notably, Clade I exhibited variation in taxa composition at X = 1.0 and X = 1.25 relative gap widths, resulting in 21 MOTUs. Specifically, at X = 1.0, sequences of *B. yothersi* from *Cecropia*

*pachystachya* (MG458858) were identified as a potential distinct species. Interestingly, at X = 1.25, specimens from *C. pachystachya* clustered within *B. yothersi*, whereas *B.* n. *yothersi* 2 from *Delonix regia* (MG458843, MG458845) formed a separate lineage from other specimens of *Citrus* sp. (MG458827, MG458830) within *B.* near *yothersi* 2. At X = 1.5, 20 MOTUs were identified, and no isolated species were detected within *B. yothersi* and *B.* near *yothersi* 2. The *p*-distance algorithm applied to the *COI* dataset returned fewer MOTUs. The putative species within *B. chilensis* did not emerge when this algorithm was used, and 18 MOTUs were identified. When X = 1.0 and X = 1.25 were used, 17 species consistent with the phylogeny/morphology and one putative species were identified. Notably, at X = 1.0, *B.* near *yothersi* 2 from *D. regia* (MG458843, MG458845) was identified as a distinct candidate species separated from those of *B.* near *yothersi* 2 from *Citrus* sp. (MG458827, MG458830). Similar results were observed when the K2P distance was applied at X = 1.25. In contrast, at X = 1.25, specimens from *D. regia* and *Citrus* sp. grouped together as *B.* near *yothersi* 2, whereas *B. yothersi* from *C. pachystachya* (MG458858) remained separate, no partitions were observed at X = 1.5.

Table 4. Number of candidate species revealed ABGD analysis of *Brevipalpus* sequences per fragment (*COI* DNA barcoding, *COI* DNF-DNR, subunit D1-D3, and ITS2 region) estimated by *p*-distance and Kimura-2-parameters substitution model algorithms

| Genome region | Model Algorithm | X | No. of species | PID | PMD | BGD |
|---|---|---|---|---|---|---|
| **COI barcoding** | K2P distance | 1.00 | 21 |  | 0.0046 | 0.016 |
| | | 1.25 | | 0.001 | 0.0077 | 0.058 |
| | | 1.50 | 20 | | 0.0046 | 0.058 |
| | *p*-distance * | 1.00 | 18 | 0.001 | 0.001 | 0.027 |
| | | 1.25 | 18 | | | 0.051 |
| **COI DNF–DNR** | K2P distance | 1.00 | 20 | 0.00278 | 0.00774 | 0.016 |
| | | 1.25 | | | | |
| | | 1.50 | | | | |
| | *p*-distance * | 1.00 | 20 | 0.001 | 0.0046 | 0.018 |
| **D1–D3 subunit** | K2P distance | 1.00 | 19 | 0.00167 | 0.00167 | 0.006 |
| | | 1.25 | | 0.001 | 0.001 | |
| | | 1.50 | | | | |
| | *p*-distance | 1.00 | 20 | 0.001 | 0.001 | 0.006 |
| | | 1.25 | | | | |
| | | 1.50 | | | | |
| **ITS2 region** | K2P distance * | 1.5 | 18 | 0.001 | 0.001 | 0.042 |
| | *p*-distance | 1.00 | 17 | 0.001 | 0.001 | 0.047 |
| | | 1.25 | | | | |
| | | 1.50 | | | | |

X = relative gap width. Nb species = number of candidate species. PID = prior intraspecific distance, partition using a range of 0.001 ($P_{min}$) to 0.1 ($P_{max}$) (0.001, 0.001668, 0.002783, 0.004642, 0.007743, 0.012915, 0.021544, 0.035938, 0.059948, 0.1000). PMD = prior maximal distances. BGD = barcode gap distance. * X = 1.0, 1.25 or 1.5 means that no partition was observed.

The **COI DNF–DNR** sequences analyzed by the **K2P** distance (X = 1.0, 1.25, and 1.5) revealed 20 MOTUs. This number of MOTUs matched the 18 species/putative species shown by phylogeny along with two candidate species: *B. papayensis* from *Citrus* sp. (DQ450486, DQ450485) and *B. obovatus* from *Cestrum nocturnum* (KC291386), both from Brazil. The latter specimens were sequenced only for this fragment. Using the ***p*-distance** model (X = 1) the same main species previously described for the K2P were recovered. However, for X = 1.25 and 1.5, the analyses resulted in an inaccurate assessment of the species with many partitions emerging as 29 putative species. This structure does not correspond to the scenario that arises from our studies based on conventional taxonomy.

Analysis of the subunit **D1-D3** sequences using the **K2P** model and the three values of relative gap width (X = 1.0, 1.25, and 1.5) revealed 19 MOTUs (**Table 4**), three more than those detected by phylogenetic analysis (16 MOTUs; **Table 2**). The distribution of sequences differed from that observed for the mitochondrial markers. For example, *Brevipalpus yothersi*, *B.* near *yothersi* 1, and *B.* near *yothersi* 2 were grouped into a single clade. Another difference was observed within *B. incognitus* and its putative species: two sequences of *B. incognitus* from *Cocus nucifera* (MK293652 - MK293653) in Brazil were distinct and formed a new putative taxon, separate from the other comprising *B. incognitus, B.* near *incognitus* 1, and *B.* near *incognitus* 2. *Brevipalpus obovatus* was assigned to three candidate lineages depending on the origin of their host plants, *i.e.*, populations from *Coniza bonariensis* (MK293694- MK293697) and *Helichrysum stoechas* (MK293700 - MK293701) from Spain, and *Solanum violifolium* (MK293698 MK293699) from Brazil. In Brazil, the sequence KP276421 of *B. papayensis* (unknown host plant) was sequenced only for this fragment and was separated from *B. papayensis* (MT664799-MT664800), which was found on *Coffea* sp. leaves. Additionally, *Brevipalpus chilensis* (Chile) from *Magnolia grandiflora* (MK293716; MK293671) was distinct from specimens collected from *Ribes punctatum* (MK293643-MK293646; MK29371-MK293715). The *p*-distance analysis (X = 1.0, 1.25, and 1.5) identified an extra taxon (20 MOTUs) with clusters similar to those observed for K2P. However, *Brevipalpus* n. *yothersi* 2 from *D. regia* (MK293681; K293683) was separate from *B. yothersi*, *B.* near *yothersi* 1, and *B.* near *yothersi* 2 from *Citrus* sp., suggesting a new candidate species.

Eighteen MOTUs were recovered by applying the K2P (X = 1.25, 1.5) for ITS2 sequences, in contrast to the 15 taxa recovered by phylogeny. *Brevipalpus yothersi* and its putative species (B. near *yothersi* 2) grouped with *B.* near *incognitus* 1, whereas *B. incognitus* and *B.* near *incognitus* 2 stayed apart. There were no ITS2 sequences obtained for the *B.* n. *yothersi* 1 specimen (**Table 2**). The partitions of the three following species differed according to the host plant or geographical region. *Brevipalpus papayensis* has been split into two different lineages according to the host plant (*Ligustrum sinense*, MH818180-MH818186 and *Coffea* sp., MT660805-MT660806). A similar partition pattern was recorded for *B. obovatus,* which had three lineages - *C. bonariensis* (MH818198-MH818201) and *H. stoechas* (MH818195-MH8181957) from Spain, and *L. sinense* from Brazil (MH818177-MH818179); and *B. chilensis* had three lineages - *M. grandiflora* (MH818206), *R. punctatum*, (MH818127 and MH818205), and *L. sinense* (MH818139-MH818142). The smallest relative gap width (X = 1.0) resulted in 66 putative species; hence it was disregarded. The *p*-distance model recovered 17 MOTUs 's (X = 1.0, 1.25, 1.5). Here, *B. incognitus* clustered with *B. yothersi*, *B.* near *yothersi* 2, and *B.* near *incognitus* 1. Only *B.* near *incognitus* 2 stayed apart in Clade I; all 16 remaining taxa split as shown by K2P.

### *ASAP-Assemble Species by Automatic Partitioning tool*

ASAP results are summarized in **Table 5** and **Figure SM1-8.** Using an adequate ASAP-score (5.5) and a high probability (1.89e-02), the *COI* barcoding sequences, processed by the K2P method, split into 17 taxa, consisting of 11 species and six putative species, keeping the same structure recovered by morphological taxonomy (**Figure SM1-8**). The *COI* barcoding dataset calculated by the *p*-distance method revealed 16 MOTUs, as *B.* near *incognitus* 1 and *B.* near *incognitus* 2 were recognized as a single group.

Analysis of *COI* DNF-DNR, subunit D1-D3, and region ITS using the ASAP method produced similar results both under the K2P and *p*-distance models. Therefore, only the results obtained under the K2P model are described further. The *COI* DNF-DNR sequences returned 20 taxa, including 11 species, 7 putative species, and 2 extra new candidates: *B. papayensis* from *Citrus* sp. (DQ450485-DQ450486) and *B. obovatus* from *Cestrum nocturnum* (KC291386), in Brazil. The ABGD method confirmed this pattern. The latter sequences (DQ450485/DQ450486 and KC291386) were unique to the *COI* DNF-DNR fragment.

Fifteen species were detected by ASAP for sequences of the subunit D1-D3. *Brevipalpus* near *yothersi* 1 and 2 were clustered together with *B. yothersi* in a single taxon. Likewise, *B.* near *incognitus* 1 was also grouped with *B. incognitus*. *Brevipalpus* n. *incognitus* 2 remained isolated.

Additionally, three candidate species were assigned to the *B. obovatus* taxon namely, *B. obovatus* from *C. bonariensis* (MK293695-MK293697) and *H. stoechas* (MK293700-MK293701) from Spain, and *S. violifolium* (MK293698-MK293699) from Brazil. This structure was revealed under the best score suggested by ASAP (score = 11.50, *p*-value = 6.61e-01).

**Table 5.** Number of *Brevipalpus* species (K2P method) predicted with the best Assemble Species by Automatic Partitioning (ASAP) score based on mitochondrial (*COI* DNA barcode and *COI* DNF-DNR) and nuclear (subunit D1-D3 and ITS2) sequences.

| Genome region | Nb of species | Partition | ASAP score | *P-val* (rank) | W (rank) | Threshold Dist. (*Dt*) | inter/intra | Distance (*Dc*) |
|---|---|---|---|---|---|---|---|---|
| ***COI* barcoding** | 17 | 3.00 | 5.5 | 1.89e-02 | 2.87e-04 | 0.018560 | 2.87e-04 | 0.018587 |
| ***COI* DNF-DNR** | 20 | 10.00 | 15.00 | 6.43e-01 | 4.94e-06 | 0.015320 | 0.016690 | 4.94e-06 |
| **D1-D3 subunit** | 15 | 7.00 | 11.50 | 6.61e-01 | 3.39e-05 | 0.008658 | 3.39e-05 | 0.009759 |
| **ITS2 region** | 17 | 3.00 | 7.50 | 2.69e-01 | 9.32e-05 | 0.029646 | 9.32e-05 | 0.33167 |

W = relative gap width; *p*-value = probability of merging the groups within the node of the ultrametric hierarchical clustering tree; *Dt* = threshold distance to which this node corresponds (jumping distance); inter/intra = threshold between intra- and interspecific distance; *Dc* = current clustering distance (grouping distance).

The ITS2 sequences revealed a structure with 17 MOTUs. Similar to the subunit D1–D3, a single taxon was recovered for *B. yothersi* and *B.* near *yothersi* 2. *Brevipalpus incognitus* and its putative species remained separate. *Brevipalpus papayensis* lineages from *L. sinense* (MH818180- MH818186) were distinct from those collected from *Coffea* sp. (MT664805-MT664806). Likewise, *B. obovatus* lineages from *C. bonariensis/H. stoechas* in Spain (MH818195-MH818197/MH818198- MH818201) have been separated from *L. sinense* in Brazil (MH818177-MH818179). *Brevipalpus chilensis* was assigned into three candidate lineages, *i.e.*, populations collected on *M. grandiflora* (MH818206), *R. punctatum* (MH818127 and MH818205), and *L. sinense* (MH818139- MH818142) (**Figure 2B**).

### *Brevipalpus* species: *Intra- and interspecific thresholds based on polynomial regression fitting*

Threshold values of the genetic distance (GD) for demarcation of *Brevipalpus* species were calculated for each region of the genome, and the results are summarized in **Table 6** (see **Figure SM1-9**). The maximum value of the confidence interval, CI-Max, represents the highest point of the polygon (CI-max = 3.95%); values higher than the (>CI-Max) value observed between two taxa support that they are different species. The magnitude and extent of the distance values varied according to the fragment analyzed. The subunit D1-D3 variation values were lower than those observed for the mitochondrial regions, with $y$ = 1.76% (0.7-2.82%) and α = 2.04%. For the ITS2 region, values extending upward were registered and $y$ = 7.75% (CI = 4.37-11.14%) and α = 6.53% (**Table 6**). The amplitude of the confidence interval was high for the ITS2 fragment (6.77) and for the *COI* DNF-DNR (4.42).

**Table 6.** Number of sequences and pairwise distance comparisons for *COI* barcoding, *COI* DNF–DNR, subunit D1–D3, and ITS2 sequences of *Brevipalpus* species using the *p*-distance model. Values were calculated using a polynomial regression fitting curve with a confidence interval of 95%.

| Genome region | Number of sequences | number of pairwise comparisons | Number of inter difference values | Intersection point (*y*) | Confidence intervals (CI) (%) | highest intraspecific vs. lowest interspecific (α) (%) |
|---|---|---|---|---|---|---|
| *COI* barcoding region | 79 | 3,487 | 136 | 2.54% | 1.14–3.95 | 3.27 |
| *COI* DNF-DNR | 427 | 91,807 | 153 | 3.34% | 1.31–5.73 | 3.23 |
| D1–D3 | 96 | 4,657 | 121 | 1.76% | 0.7–2.82 | 2.04 |
| ITS2 | 117 | 6,787 | 153 | 7.75% | 4.37–11.14 | 6.53 |

## DISCUSSION

### Species-group rearrangements and informative morphological traits

The multilocus and integrative approach adopted in this study allowed for substantial advances in species-group rearrangements as well as species-delimitation criteria. Although only 11 valid species were evaluated in this study out of the more than 300 valid species (Castro et al., 2025), the analysis included taxa representing four of the six species groups established by Baker & Tuttle (1987) and all species currently of major agricultural importance (Hebert et al., 2003; Childers & Rodrigues, 2011; Beard et al., 2015). Our results showed that the current division of *Brevipalpus* into species groups only partially corresponded to phylogenetic branches; therefore, a rearrangement of species groups is proposed (**Table 7**). The findings allow a discussion on phylogenetically informative morphological traits and a review of the definition of each species group.

**Table 7** summarizes the morphological characters used to separate the currently accepted six species groups according to Baker & Tuttle (1987), and highlights the discrepancies between these criteria and the molecular analyses herein obtained. The phylogenetic study has uncovered the following inconsistencies, mostly due to homoplasy in the morphological characters formerly used to define species groups: (i) the *phoenicis* species-group, that included *B. yothersi*, *B. incognitus*, *B. phoenicis*, *B. papayensis*, and *B. ferraguti*, is polyphyletic and the species were set into two distinct clades (Clades I and IV). Remarkably, *B. yothersi* and *B. papayensis*, two former synonyms of *B. phoenicis* resurrected by Beard *et al.* (2015), have been placed in two different clades. Additionally, *B. phoenicis*, the exemplar species of the *phoenicis* group, was clustered together with *B. obovatus*, the exemplar species of the *obovatus* group, in Clade IV; (ii) *Brevipalpus lewisi* and *B. oleae* considered to be part of the *cuneatus* species-group, along with two potential species displaying morphological traits of the *cuneatus* species-group, were also found to be polyphyletic and were split into three clades (Clades III, V, and VI). *Brevipalpus lewisi* was clustered with *B. californicus ss*, the exemplar species of the *californicus* group, in Clade III. *Brevipalpus oleae* formed a separate Clade (Clade VI) distinct from *B. mallorquensis* and *B.* n. *mallorquensis* (Clade V), as previously noted by Alves *et al.* (2019). Interestingly, the new species from *Citrus*, Argentina (MG458828/29; MH204695/96; MK293669/70; MH818133/34) were separated as a different lineage (Clade II), although following Baker and Tuttle they should be placed in the *obovatus* group.

**Table 7.** Arrangement of *Brevipalpus* taxa according to the molecular-based clades, main morphological traits, and their membership in the species-group established by Baker & Tuttle (1987).

| Clade | species & putative species | DLO setae | f2 | T2 sol. | Palp seg* | palp setae* | Baker & Tuttle (1987) (BP) |
|---|---|---|---|---|---|---|---|
| I | *B. yothersi* | 6 | 0 | 2 | 4 | 3 | *phoenicis* |
| | *B.* near *yothersi* 1 | 6 | 0 | 2 | 4 | 3 | *phoenicis* |
| | *B.* near *yothersi* 2 | 6 | 0 | 2 | 4 | 3 | *phoenicis* |
| | *B. incognitus* | 6 | 0 | 2 | 4 | 3 | *phoenicis* |
| | *B.* near *incognitus* | 6 | 0 | 2 | 4 | 3 | *phoenicis* |
| II | B. n. sp. Argentina | 6 | 0 | 1 | 4 | 3 | *obovatus* |
| III | *B. californicus* | 7 | 1 | 2 | 4 | 3 | *californicus* |
| | *B.* near *californicus* 1 | 7 | 1 | 2 | 4 | 3 | *californicus* |
| | *B.* near *californicus* 2 | 7 | 1 | 2 | 4 | 3 | *californicus* |
| | *B. lewisi* | 7 | 1 | 1 | 4 | 3 | *cuneatus* |
| IV | *B. chilensis* | 6 | 0 | 1 | 4 | 3 | *obovatus* |
| | *B. obovatus* | 6 | 0 | 1 | 4 | 3 | *obovatus* |
| | *B. ferraguti* | 6 | 0 | 2 | 4 | 3 | *phoenicis* |
| | *B. papayensis* | 6 | 0 | 2 | 4 | 3 | *phoenicis* |
| | *B. phoenicis* | 6 | 0 | 2 | 4 | 3 | *phoenicis* |
| V | *B. mallorquensis* / *B.* near *mallorquensis* | 7 | 1 | 1 | 4 | 3 | *cuneatus* |
| VI | *B. oleae* | 7 | 1 | 1 | 4 | 3 | *cuneatus* |

**DLO setae** = number of dorsolateral setae on opisthosoma; *f2* = dorsal setae (presence = 1 or absence = 0); **T2 sol.** = number of solenidia at the apex of tarsus II of the female; **palp seg.** = number of papal segments; **palp setae** = number of setae on the apical segment of the palpus; **BP** = groups following Baker & Tuttle 1987 indicating where the species in this study would be classified if following Baker & Tuttle (1987) classification. * Non-variable traits are shown for taxonomic context and comparison with future studies

Considering our results, a new species-group rearrangement is proposed. To avoid confusion with all the previous *Brevipalpus* classifications (Baker et al., 1975; Meyer, 1979; Baker & Tuttle, 1987), in which specific names were used to designate species groups, a Roman numeral nomenclature is adopted. As the taxonomic knowledge of the genus evolves and new genetic lineages are detected, subsequent Roman numerals may be assigned. Hence, six *Brevipalpus* species-groups are here proposed, currently composed of the following species: I) *B. yothersi* and *B. incognitus*; II) *B.* new sp. collected from *Citrus* in Argentina; III) *B. californicus* and *B. lewisi*; IV) *B. papayensis, B. phoenicis*, *B. obovatus*, *B. ferraguti*, and *B. chilensis*; V) *B. mallorquensis* collected from *Rosmarinus officinalis*, and a species near *B. mallorquensis* from *Helichrysum stoechas*, both from Spain; and VI) *B. oleae* (**Table 7**). Additionally, the cryptic species detected in this study are variously members of *Brevipalpus* species-groups I, III and V. Although some of the proposed species-groups are currently composed of only one species, such groups are expected to increase in size as new taxa are characterized and included.

Incongruences between the phylogenetic groups and the Baker & Tuttle classification indicate that some of the accepted taxonomic criteria do not represent the affinity relationships between species. Species clustered in the same Clade share the same state for some of these characters but not for others. The number of dorsolateral setae on the opisthosoma was consistent among species of the same Clade (**Tables 7; SM2-20**) and, according to current knowledge, could be useful as a species-group diagnostic character. However, distantly related clades, *e.g.*, III and V/VI, shared the presence of the seta *f2* totaling seven dorsolateral setae, suggesting that this trait arose independently in separate lineages in the genus and would not be phylogenetically informative. In contrast, the number of solenidia on tarsus II varied among species in clades III and IV (**Table 7; SM2-20**), indicating that this feature likewise exhibits homoplasy and should not be used for division of the genus, supporting what many authors have already noted (Groot et al; 2005; 2006 (Nascimento et al. 2024).

Unfortunately, it was impossible to evaluate the phylogenetic value of the number of palpal segments and the number of setae on the distal segment of the palps from the analyzed dataset. These traits were used to delimit the *portalis* and *frankeniae* species-groups, which were not represented in this study, and were not variable among species that were studied here. However, discrepancies in the morphology of the *portalis* species-group have already been pointed out by Alves *et al.* (2019), and they classified a species morphologically very similar to *B. cuneatus*, *B. sulcatus* Alves, Ferragut & Navia, in *portalis* species group (Alves et al., 2019). A better understanding of the phylogenetic relevance of these characters depends on the incorporation of species with variable traits into the *Brevipalpus* phylogeny, including those previously classified as belonging to the *portalis* and *frankeniae* groups.

Among the morphological characters recently considered in the taxonomy of *Brevipalpus* (Navia et al., 2013; Beard et al., 2015; Alves et al., 2019) and described in **Table SM2-20, Figures 4A, 4B and SM1-10**, we conclude that the cuticular patterns of the dorsal and ventral shields, microplate ornamentation (**Figure SM1-11),** and the shape of the palp femurogenu seta are species-specific. These characteristics vary among species within the same phylogenetic group, and therefore are not phylogenetically informative at the species-group level.

**Spermatheca character and its implications for the recognition of groups in *Brevipalpus***

Tree analyses supported the spermathecal patterns as a reliable and suitable criterion for distinguishing species-groups (**Figures 4A and 4B**, **SM1-10** and **12**; **Table SM2-20**). The shape of the seminal vesicle was homologous among closely related taxa within the same Clade (**Figure 4B**; **SM1-10** and **12**). The spermatheca or seminal receptacle is a part of the female insemination system whose function is to store and protect the sperm until its use in egg fertilization (Alberti & Kitajima, 2014). The importance of the spermatheca in the taxonomy of *Brevipalpus* and other tenuipalpids has been historically largely ignored. Castagnoli (1974) was the first to describe this organ in detail for in eight flat mite species, although she did not mention its taxonomic importance. Later, Baker & Tuttle (1987) briefly described and illustrated the spermatheca in some tenuipalpids from Mexico, including eight species in *Brevipalpus*, without using it for species delimitation. Recent studies (Beard et al., 2015; Di Palma et al., 2020) have highlighted the spermatheca as a key morphological character in species descriptions (Navia et al., 2013; Beard et al., 2015; Alves et al., 2019). In other plant mites, the spermatheca has long been recognized as a valuable taxonomic trait, including representatives of the spider mite family Tetranychidae (van Eyndhoven & Vacante, 1985; Seeman et al. 2017) belonging, like the flat mites, to the superfamily Tetranychoidea and, therefore, related to them.

The findings of this study show that *Brevipalpus* species included in the species Group I have an oval, thin-walled spermatheca with a smooth outer surface and a long (at least twice as long as the vesicle), thick stipe on the distal pole (**Figure SM1-10A** to **10F; SM1-12** and **13**); the single species in Group II has a rounded, thin-walled spermatheca, with a crown of projections on the distal pole (**Figure SM1-10G**); species in Group III have a somewhat flattened subspherical spermatheca, the distal margin notably thick-walled to create a reduced inner space, with a crown of projections on both the distal and proximal poles (**Figure SM1-10H** to **10K**); species in Group IV have a rounded to slightly flattened spermatheca, the distal margin thin-walled with a large internal space, projections over on the entire surface, distal projections tend to be shorter and directed distally, proximal projections tend to be longer, finger-like, and directed proximally (**Figure SM1-10L** to **10P**); species in the Group V have an elongate inverted piriform spermatheca, with the narrower elongate proximal region covered with fine projections in the proximal half (**Figure SM1-10Q** and **10R**); and species in Group VI have a similar inverted piriform spermatheca to that of Group V, but the narrower proximal region is much shorter with fine projections (**Figure SM1-10S**). Although there is a high correlation between the shape of the female spermatheca and the genetic grouping, to avoid further confusion, we recommend that the morphology of this structure should not be used in isolation for species-group identification, but in combination with other reliable traits.

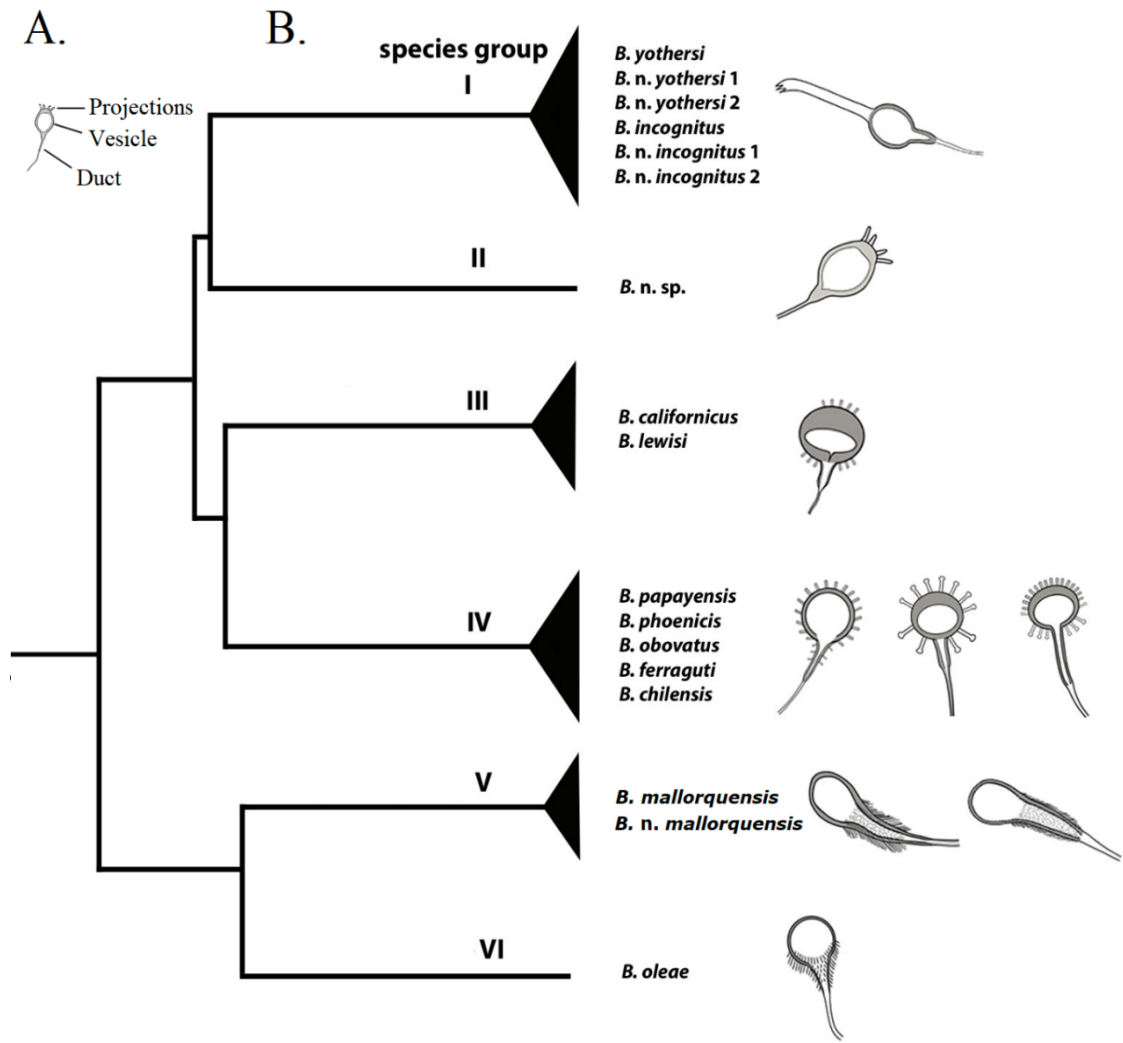

**Figure 4. A.** Spermathecal morphology, including a duct, a vesicle, and projections or ornamentation in the distal end. **B.** Tree representation showing the phylogenetic relationships among *Brevipalpus* species and candidate species based on concatenated datasets (*COI* barcoding, *COI* DNF-DNR, subunit D1–D3, and ITS2) and the morphology of the spermathecal vesicle is shown on the right for each Clade.

**Cryptic diversity unveiled by multiple approaches: Incongruences suggest ongoing speciation**

Phylogenies and all species delimitation methods supported the presence of cryptic diversity in certain taxa (**Figure 3**). *Brevipalpus* n. *mallorquensis* was confirmed to be a separate species by all species-delimitation methods used, supporting the findings of Alves *et al.* (2019). The DNA sequences of the specimens identified as *B. californicus s.l.* also suggest the existence of cryptic species within the taxon. This information corroborates the findings of morphological studies by Beard *et al.* (2012), who identified four morphotypes within the *B. californicus*, namely, *B. californicus s.l.*, *B. californicus s.s.*, *B. californicus* species B, and *B. californicus* species C. The congruences among multiple species-delimitation methods used in this study support the hypothesis of new taxa, the description of which are in preparation.

However, for some taxa, particularly in Clades I and IV, there were uncertainties between the species-delimitation methods (**Figure 3**). Although the phylogenetic and concatenated trees, both of single genetic markers and concatenated, showed strong clustering at the *Brevipalpus* species level, some putative species could not be differentiated through phylogenetic analyses (BPP ≥ 98%) or by some of the genetic distance techniques. Generally, there was an overestimation of the number of lineages when using delimitation approaches other than phylogeny and formal morphology (see **Figure 3, Table SM2-20**).

The discordance between the mitochondrial and nuclear markers could be attributed to a multitude of reasons, including incomplete lineage sorting, sex-biased dispersal, asymmetrical introgression, natural selection, or genetic sweeps mediated by cytoplasmic bacteria (Després, 2019), such as *Cardinium* (Kitajima et al., 2007). Additionally, low mitochondrial diversity was observed in populations with high infection rates of symbiotic bacteria like *Wolbachia* (Deng et al., 2021), which could have affected the results of the delimitation methods. The various transmission modes of mitochondrial and nuclear markers (maternal versus biparental) could explain their differences in sensitivity in the population analysis over time. Mitochondrial DNA is maternally inherited and does not undergo recombination, thus it represents a single lineage, whereas nuclear DNA is inherited from both parents, undergoes recombination, and could mix genetic material, potentially obscuring lineage boundaries (Després, 2019). The transplantation experiments conducted with *B. phoenicis* on various host plants suggest that the clonal adaptation of these mites could be explained by the frozen niche variation model (Groot et al., 2005). This model proposes that clones specialized in various niches best explain the clonal adaptation of these mites. The lineages within *B. yothersi* (detected only by *COI* sequences), similar to those from *Cecropia* (Brazil), and within *B.* n. *yothersi* 2, such as *Delonix and Citrus* (Brazil and Argentina), could be explained by frozen niche variation.

These uncertainties, or lack of effective congruence suggest that some speciation processes are still in progress and that this study "captured" distinct phases of these processes, such as those designated as *B.* near *yothersi* and *B.* near *incognitus* in Group I. Limited knowledge of the ecological traits, distribution, host range, and reproduction mode of *Brevipalpus* mites hinders a detailed discussion of the drivers and mechanisms of speciation events. Despite this, some hypotheses could be proposed.

Results of species delimitation in populations morphologically identified as *B. incognitus* suggest a nonadaptive radiation process, indicating lineage diversification without significant ecological divergence (*e.g.*, same host plant and geographic overlap). *Brevipalpus incognitus* was originally described from coconut trees in southeastern Brazil (Navia et al., 2013) and has not yet been reported in other countries. Subsequently, it was also found on *Annona muricata* and ornamental *Phoenix* sp. palms in the northern and southeastern regions of the country, respectively. Surprisingly, specimens collected from the type host and locality for *B. incognitus* and identified as *B. incognitus* consisted of different phylogenetic lineages, *i.e.*, *B.* near *incognitus* 1 and 2. These genetic lineages seem not to exhibit ecological differences. Although this is the clearest example of the occurrence of nonadaptive radiation among the putative species studied, it could be a common process for *Brevipalpus* mites. Adaptive radiation has long been considered the standard evolutionary pattern (Losos & Mahler, 2010), but nonadaptive radiations are also common and are genuine phenomena related to the properties of organisms as well as to ecological conditions (Czekanski-Moir & Rundell, 2019). Nonadaptive radiation generally involves the proliferation of species as a result of the restriction of gene flow (Czekanski-Moir & Rundell, 2019). Limited information is available on the reproductive behavior of *Brevipalpus* mites, as only a few economically important species have been studied. In species such as *B. chilensis* (Group IV), the female to male ratio is nearly 1:1 and sexual reproduction is the norm (Alberti et al., 2014) and there is no observed presence of *Cardinium* (Quiros-Gonzalez, 1986). Males are prevalent in species from the *cuneatus* (Group V) and *portalis* groups, which combined represent most of the diversity in the genus, whereas males tend to be rare in other species-groups.

One possible reason for reproductive isolation is the cytoplasmic incompatibility (CI) caused by symbionts. Infections by *Cardinium* and *Wolbachia* have caused CI in numerous arthropods (Shropshire et al., 2020), including mites (Gotoh et al., 2007; Cruz et al., 2021). In addition to *Cardinium*, *Wolbachia* has been found in the *Brevipalpus* bacteriome (Ospina et al., 2016). Although *Cardinium* does not cause CI in *B. yothersi*, it is possible that variants infesting other species could cause it. Therefore, CI induced by symbionts could likely drive speciation for some *Brevipalpus* mites, as has been demonstrated for other arthropods (Bordenstein et al., 2001; Jaenike et al., 2006). Additionally, other mechanisms of reproductive isolation between haplodiploid spider mite populations (family Tetranychidae) are under investigation (Cruz *et al.*, 2021; Engelstädter & Hurst, 2009; Knegt et al., 2016; Cruz et al., 2021) and may provide clues to analyze similar processes in *Brevipalpus*. Studies comparing the importance of the mode of reproduction linked to the phylogenies are rare. A comparison of sexual monogononta rotifers and obligately asexual bdelloids (Tang et al., 2014) is consistent with the hypothesis that bdelloids diversified at a faster rate into less discrete species because their diversification does not depend on the evolution of reproductive isolation. This highlights other points to explore for *Brevipalpus* populations.

The results also suggested cryptic speciation in closely related populations of *B. yothersi* associated with citrus in Argentina and Brazil. In this case, adaptative radiation may have occurred when populations became physically isolated from each other for a long time or due to underlying divergent selection for adaptation to different niches (adaptative radiation), as reported for parthenogenetic oribatid mites (Birky & Barraclough, 2009; Heethoff et al., 2009). Although the new genetic lineages can be phenotypically distinct in these soil-dwelling oribatid mites, some cases of cryptic speciation detected only by genetic analysis have also been reported (Birky & Barraclough, 2009). Maraun *et al.* (2003) observed that the genetic differences between closely related parthenogenetic species were larger than those between closely related sexual species, indicating that parthenogenetic lineages may radiate slower than sexual species.

**Exploring the species-delimitation methods and molecular markers**

The findings confirmed the presence of a new species in Clade II (*B.* new sp.) and the status of Clade VI (*B. oleae*) through consistent and well-supported hypotheses, achieved by morphology, phylogenic inferences, and distance delimitation methods (**Figure 3**). Putative species in Clades III and V, B. n. *californicus* 1, B. n. *californicus* 2, and B. n. *mallorquensis*, were also supported by phylogeny and by all species-delimitation methods. However, ambiguities were observed for some taxa in Clades I and IV (**Figure 3**).

In Clade I, a distinct lineage of *B.* n. *yothersi* 2 from *Delonix regia* (Piracicaba, BR) was identified, separate from the *B.* n. *yothersi* 2 collected from *Citrus* sp. (Argentina) based on *COI* fragments, whereas the results for the nuclear fragments were quite variable. Similarly, *B. yothersi* from *Cecropia pachystachya* was regarded as an isolated lineage only by the *COI* barcode fragment. Unlike mitochondrial markers, the nuclear markers did not detect enough nucleotide divergence to distinguish all the putative species in Clade I. In Clade IV, potential cryptic species were identified within *B. papayensis*, *B. obovatus*, and *B. chilensis*, each distinctly associated with a specific host plant (**Table SM2-19**; **Figure 3**) and with consistent results supporting this both molecular markers and methods. *Brevipapus papayensis* from *Citrus* sp. (*COI* fragment) and from *Coffea* sp. (nuclear fragments) were separated from specimens collected from other hosts. Additionally, *B. obovatus* from each of *C. bonariensis*, *H. stoechas* and *S. violifolium* and *B. chilensis* from each of *M. grandiflora*, *R. punctatum* and *L. sinense* (**Table SM2-19**) were separated using the ABGD and ASAP approaches based on mitochondrial and nuclear markers (**Figure 3; SM2-19**).

The ABGD results were dependent on the chosen model (Kimura and *p*-distance) and were also affected by changes in the gap width parameter. Applying the standard settings (X = 1.5), no barcode gap was observed, and only one partition was detected for the mitochondrial fragments. When lowering the X-value (1.0; 1.25), the *COI* sequences were clustered into different entities, which was more congruent with the phylogeny and morphology findings. Similar results were observed by Jörger *et al.* (2012) regarding sea slugs. Species recovered based on the *COI* DNF-DNR dataset were only distinguished when a smaller relative gap width (X = 1) was used, as larger values (1.25 and 1.5) grouped all sequences together. Therefore, for analyzing the *COI* DNF-DNR, an X = 1 was sufficient to detect potential partitions among the *Brevipalpus* populations. Similarly, smaller relative gaps in the *COI* barcode dataset revealed distinct species, whereas larger gaps grouped all sequences into a single taxon. The k2P model yielded more entities than the *p*-distance models across various gap widths, except for the nuclear marker D1-D3. The barcoding gaps tend to be larger under the K2P model than with uncorrected *p*-distance models (Srivathsan & Meier, 2012). For sequences with small interspecific distances (<5%) from closely related species, *p*-distances were deemed more suitable than K2P (Nei & Kumar, 2000). Previous studies have shown that using *p*-distances could result in higher or similar identification success rates than K2P correction (Srivathsan & Meier, 2012).

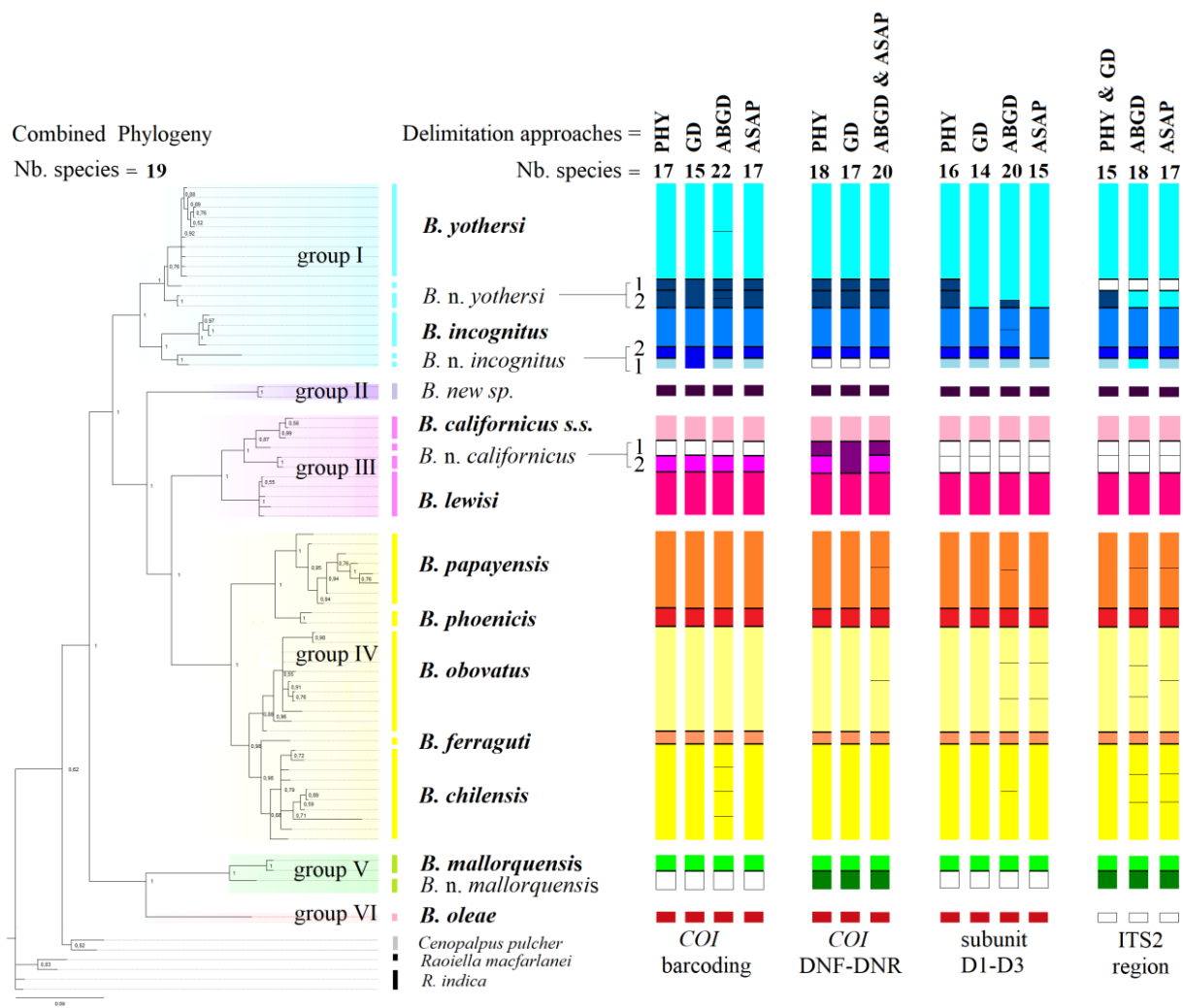

**Figure 3.** Bayesian inference tree showing the phylogenetic relationships among *Brevipalpus* lineages based on concatenated datasets (*COI* barcoding, *COI* DNF-DNR, subunit D1–D3, and ITS2). Differently colored clades represent species groups, with bars of the same color tones on the right indicating the number of species (bold) and putative species (regular) according to phylogeny (PHY), genetic distance (GD), and delimitation approaches ABGD and ASAP using K2P and *p*-distance models, various partitions, and relative gap width values (X = 1.0, 1.25, 1.5). White samples are not available. Bayesian posterior probability (BPP) values ≥0.98 were considered to be indicate well-supported putative species.

**DNA barcoding for *Brevipalpus***

In this study, the distance values corroborated some of the putative species but were variable according to the gene fragment or region used. Some *Brevipalpus* species have shown maximum intraspecific divergences equal to or greater than the minimum and mean interspecific distances, indicating an overlap between the maximum intra- and minimum interspecific values. The highest intraspecific divergence (K-2P) in *COI* barcoding was observed within *B. yothersi* sequences (2.44%), surpassing the mean and minimum divergence between *B.* n. *yothersi* 1 and *B.* n. *yothersi* 2 ($\underline{X}\ distance\ = 2.22\%$, 1.99%–2.44%) and between *B.* n. *incognitus* 1 and *B.* n. *incognitus* 2 ($\underline{X}\ distance\ = 1.99\%$, 1.99%–1.99%). There was no overlap of these values in the other species and putative species. In the *COI* DNF-DNR fragment, the highest intraspecific variability occurred within *B. yothersi* (4.28%) and *B. papayensis* (3.48%). These values exceeded most of the minimum interspecific distances and some of the mean distances obtained for species grouped in Clade IV. Some closely related species (*e.g.*, *B. yothersi* versus *B. incognitus*, *B. chilensis* x *B.*

*obovatus*) showed extensive amounts of intraspecific variation, often exceeding the interspecific variation between the species (**Table 3**; **SM2-11–SM2-18**). Therefore, the general barcoding assumption that intraspecific variation is smaller than interspecific variation was disregarded for some *Brevipalpus* species, as previously reported in the genus *Tetranychus* (Ros & Breeuwer, 2007).

**The boundaries of the species delimitation**

Establishing GD thresholds within and between species is important for molecular diagnostics of *Brevipalpus* species (**Table 6**). Based on the *COI* barcoding sequences, the lowest GDs between the trusted valid species *B. chilensis* vs. *B. ferraguti* and *B. chilensis* vs. *B. obovatus* were 2.68% and 2.90%, respectively, which were above $y$ = 2.54%. These species are morphologically distinct, and there is no doubt about their identification, thus, $α$ = 3.27% could be a reliable value to differentiate species. The selection of the CI-max value (3.95%) enhances the security of confirming genetic distinctiveness of species. The *COI* DNF–DNR showed $y$ = 3.34 (1.31%–5.73%). Above this intersection point ($y$), only valid species were observed. However, the intraspecific variability within *B. yothersi* and *B. papayensis* was high, at 4.28% and 3.48%, respectively. In studies based solely on DNA sequences, values between $α$ (3.27%) and CI-max (5.73%) could also indicate genetic differentiation of *Brevipalpus* species. The variability values for subunit D1–D3 were lower than those for the mitochondrial regions, with $y$ = 1.76% (0.7%–2.82%). The shortest genetic distances separating two valid species were 0.40% (*B. yothersi* vs. *B. incognitus*) and 1.2% (*B. papayensis* vs. *B. phoenicis*). Likewise, the $α$ = 2.07% and CI-max = 2.82% for the subunit D1–D3 indicates a reliable differentiation between species. The values for the ITS2 region were higher, with $y$ = 7.75% (4.37–11.14%). Although closely related species showed a low ITS2 threshold value, *i.e.*, *B. californicus s.s.* versus *B. lewisi* (0.93%), high intraspecific distances were recorded in *B. papayensis* (4.83%, 0–13.07%), *B. obovatus* (6.90%, 0–12.51%), and *B. chilensis* (3.73%, 0%–11.43%), causing the estimated value for $y$ to shift upward. This supports the presence of distinct lineages within these taxa as detected by ABDG and ASAP. Based on these findings, a threshold value up to $y$ = 7.75% and 11.14% (CI-max) could be used for ITS2 molecular diagnostic studies in the absence of morphological data (**Table 6**). The description of the new valid species will lower these threshold values.

Findings of previous studies on plant mites provided reference values for comparison with the findings of this study. Two closely related species of phytophagous spider mites, *Tetranychus urticae* and *T. turkestani* (Ugarov & Nikolski), exhibited significant nucleotide diversity in the *COI* region (mtDNA) (3%–4%) and minimal variation in the ITS2 region (<0.5%) (Navajas et al., 1998). Intraspecific values for single species were as follows: *Tetranychus evansi* (Baker & Pritchard) populations showed *COI* diversity ranging from 0.03% to 1.4% (Boubou et al., 2011); *Eotetranychus carpini* (Oudemans) populations presented ITS nucleotide diversity of 0.6% (Malagnini et al., 2012); *Aceria guerreronis* (Keifer) from different geographic regions worldwide exhibited ITS nucleotide diversity ranging from 0.49% to 1.90% (Navia et al., 2005); the predatory mite *Euseius nicholsi* (Ehara & Lee) had *COI* nucleotide diversity ranging from 0.342% to 1.11% and ITS nucleotide diversity ranging from 0.0% to 0.19% (Yang et al., 2012). High interspecific variation in the ITS region has been reported in the following Phytoseiidae species: *Typhlodromus phialatus* Athias-Henriot/*T. pyri* Scheuten (6.7%) (Tixier et al., 2006); *Amblyseius largoensis* (Muma)/*A. herbicolus* (Chant) (5.89%) (Navia et al., 2014), and *Neoseiulella aceri* (Collyer)/*N. litoralis* (Swirski & Amitai) (6.0%) (Kanouh et al., 2010). The results for *Brevipalpus* were in line with previously reported values for mitochondrial markers, and this study revealed new values for the subunit D1–D3. The ITS2 sequences showed high variability among the species examined. The proposed threshold values for *Brevipalpus* species streamline species assignment by linking sequence clusters with taxa, thereby facilitating taxonomic studies in cases of uncertain classifications.

## Final considerations

This study provided, for the first time, a phylogenetic background that allowed revisiting the species-group concept in the genus *Brevipalpus*, previously based on just a few morphological characters and with a practical identification purpose. Multi-locus genetic distances were analyzed through a variety of approaches to identify cryptic lineages, support species diagnosis, and highlight priority study needs. The speciation revealed in two *Brevipalpus* species-groups (I and IV), not linked to ecological differences, makes this group of mites an excellent model for evolutionary studies focusing on reproductive modes and endosymbionts.

The importance of an integrative approach to advancing the systematics of *Brevipalpus* mites is emphasized by the results of this study. It was noted that nuclear markers were essential for identifying species lineages and detecting host-associated lineages when mitochondrial DNA did not exhibit enough diversity. However, nuclear markers were unable to detect cryptic diversity within *B. yothersi* and *B. incognitus*, that was strongly suggested by the mitochondrial fragments. Additionally, both nuclear and mitochondrial markers were crucial and complemented each other in identifying candidate hosts-liked lineages within *B. chilensis* and *B. obovatus* when morphological evidence was lacking. Therefore, it is important to carefully consider which markers and models to use in species delimitation before asserting the occurrence of new cryptic species (Després, 2019), and note that morphological investigations remain crucial for evaluating genetic differences in potential new species.

The inclusion of new species, populations, and the enrichment of sequence datasets will certainly validate the threshold values given in this study. In any case, the ranges established are already quite reliable and can be used to support routine taxonomic identifications, such as pinpointing cryptic lineages, identifying specimens intercepted during transit of plant material, or for field surveys to detect vector species. It is advisable to consider all of these findings within an integrative taxonomy framework.

## Legal requirements

DN obtained a permanent license, Permit Number 20650-1, from ICMBio (Chico Mendes Institute for the Conservation of Biodiversity, Brazilian Ministry for Environment) for collecting zoological specimens in Brazil.

## Appendices

Revisiting the systematics of *Brevipalpus* mites_Supplementary Material 1, Figures -SM1
https://hal.science/hal-04907421v2

Revisiting the systematics of *Brevipalpus* mites_Supplementary Material 2, Tables -SM2
https://hal.science/hal-04907421v2

## Acknowledgements

We offer special thanks Prof. Dr. Manoel Guedes Correa Gondim Jr., Dra. Rosa de Belém das Neves Alves, and Roberto C. Trincado for their help with sample collection. We also appreciate the contributions of Prof. Dr. Elliot Watanabe from Escola Superior de Agricultura "Luiz de Queiroz", ESALQ-USP, Piracicaba, São Paulo, for his assistance in collecting specimens for the study and for supporting the scanning electron microscopy studies. We thank Dr. Gregory Evans (APHIS-USDA) for his valuable input and expertise in revising the manuscript. We thank EMBRAPA, the University of São Paulo, the Smithsonian National Museum of Natural History, and the National Agricultural Library (NAL-USDA) for their support and assistance with references. Mention of trade names or commercial products in this publication is solely for the purpose of providing specific information and does not imply recommendation or endorsement by the USDA; USDA is an equal opportunity provider and employer. We also thank Felix Sperling and the reviewers Marla Schwarzfeld, Owen Seeman and 1 anonymous for their helpful comments on a previous version of this paper. Preprint version 2 of this article has been peer-reviewed and recommended by Peer Community in Zoology, 100359, https://doi.org/10.24072/pci.zool.100359; (Sperling, 2025).

**Data, scripts, code, and supplementary information availability**

Data are available online:Mitochondrial (COI barcoding & COI DNF/DNR) and nuclear (D1-D3 28S & ITS2) DNA datasets: Revisiting the systematics of *Brevipalpus* mites https://doi.org/10.5281/zenodo.14722287

## Funding

The Brazilian Coordination for the Improvement of Higher Education Personnel (CAPES), National Program of PostDoctoral in Agronomy, Brazil for providing financial support for RSM (CAPES-PNPD/Agronomy Proc. No. 88882.305808/2018-1) and National Council for Scientific and Technological Development (CNPQ) (Proc No. 161389/2011-2). The project 'Genomics and transcriptomic of virus-vector-host plant relationship in the pathosystem of *Brevipalpus* transmitted viruses; systematic and evolution of *Brevipalpus* mites and their endosymbionts; new strategies to manage citrus leprosis in the São Paulo state (FAPESP, the São Paulo Research Foundation, Proc No. 2019/25078-9).

## Conflict of interest disclosure

The authors declare that they comply with the PCI rule of having no financial conflicts of interest in relation to the content of the article.